\documentclass[structabstract]{aa} 
\usepackage{graphicx}
\usepackage{upgreek}
\usepackage{amsmath}

\usepackage{natbib}
\usepackage{twoopt}
\usepackage{multirow}
\usepackage{caption}
\usepackage{siunitx}
\usepackage{gensymb}
\usepackage{threeparttablex}
\usepackage{subfig}

\usepackage{txfonts}
\usepackage{color}
\usepackage{hyperref}
\hypersetup{%
  colorlinks=true,
  linkcolor=blue,
  citecolor=blue
}

\def \bmath #1 {{\hbox{\boldmath{$#1$}\unboldmath}}}

\usepackage[dvipsnames]{xcolor}

\begin{document}

\bibliographystyle{aa}

\title{Gas phase Elemental abundances in Molecular cloudS (GEMS)}
\subtitle{VIII. Unlocking the CS chemistry: the CH + S$\rightarrow$ CS + H and  C$_2$ + S$\rightarrow$ CS + C reactions}
\authorrunning{Rocha et al}
\titlerunning{CS formation}
 
\author{
  Carlos M. R.  Rocha\inst{1}
  \and
  Octavio Roncero\inst{2}
  \and
  Niyazi Bulut\inst{3}
  \and
  Piotr Zuchowski\inst{4}
  \and
  David Navarro-Almaida\inst{5}
  \and
 Asunci\'on Fuente\inst{6}
 \and
Valentine Wakelam\inst{7}
\and
Jean-Christophe Loison\inst{8}
\and
Evelyne Roueff\inst{9}
\and
Javier R. Goicoechea\inst{2}
  \and
 Gisela Esplugues\inst{6}
 \and
 Leire Beitia-Antero\inst{6}
\and
Paola Caselli\inst{10}
\and
Valerio Lattanzi\inst{10}
    \and
Jaime Pineda\inst{10}
    \and
Romane Le Gal\inst{11,12}
\and
Marina Rodr\'{\i}guez-Baras\inst{6}
  \and
 Pablo Riviere-Marichalar\inst{6}
}

\institute{
          Laboratory for Astrophysics, Leiden Observatory, Leiden University, PO Box 9513, 2300-RA, Leiden, The Netherlands
          \and
          Instituto de F{\'\i}sica Fundamental (IFF-CSIC), C.S.I.C.,                                   
          Serrano 123, 28006 Madrid, Spain
          \and
       University of Firat, Department of Physics, 23169
          Elazig, Turkey
          \and
      Institute of Physics, Faculty of Physics, Astronomy and Informatics, Nicolaus Copernicus University in Torun, 
Grudziadzka 5, 87-100 Torun, Poland
            \and
   Universit\'e Paris-Saclay, CEA, AIM, D\`epartement d'Astrophysique (DAp), F-91191 Gif-sur-Yvette, France 
           \and
          Observatorio Astron\'omico Nacional (IGN), c/ Alfonso XII 3, 28014 Madrid, Spain.
        \and
        Laboratoire d'astrophysique de Bordeaux, Univ. Bordeaux, CNRS, B18N,
        all\'ee Geoffroy Saint-Hilaire, 33615 Pessac, France
          \and
          Institut des Sciences Mol\'eculaires (ISM), CNRS, Univ. Bordeaux, 351 cours de la
            Lib\'eration, F-33400, Talence, France
          \and
          Sorbonne Universit\'e, Observatoire de Paris, Univerit\'e PSL, CNRS, LERMA, 92190 Meudon, France
       \and
     Centre for Astrochemical Studies, Max-Planck-Institute for Extraterrestrial Physics,Giessenbachstrasse 1, 85748, Garching, Germany
       \and
 Institut de Plan\'etologie et d'Astrophysique de Grenoble (IPAG), Universit\'e Grenoble Alpes, CNRS, F-38000 Grenoble, France
         \and
   Institut de Radioastronomie Millim\'etrique (IRAM), 300 Rue de la Piscine, F-38406 Saint-Martin d'H\`{e}res, France
      }

 \abstract 
 {Carbon monosulphide (CS) is among a few  sulphur-bearing species that has been widely observed in all environments, including the most extreme ones such 
 as diffuse clouds.
   Moreover, it has been widely used as a tracer of the gas density in the interstellar medium in our Galaxy and external galaxies.
   Therefore, the full understanding of its chemistry in all environments is of paramount importance for the study of the interstellar matter. }
 {Our group is revising the rates of the main formation and destruction mechanisms of CS. In particular,
   we focus on those which involve open-shell species for which the classical capture model might not be accurate enough.
   In this paper, we revise the rates of reactions CH + S $\rightarrow$ CS + H and C$_2$ + S $\rightarrow$ CS + C.
   These reactions are important CS formation routes in some environments such as dark and diffuse warm gas.}
   {We performed ab initio calculations to characterize the main features of all the electronic states correlating
     to the open shell reactants. For CH+S we have calculated the full potential energy surfaces (PES)
     for the lowest doublet states and 
     the reaction rate constant with a quasi-classical method. For C$_2$+S, the reaction can only take place
     through the three lower triplet states, which all present
     deep insertion wells. A detailed study of the long-range interactions for these triplet states allowed
     to apply a statistic adiabatic method to determine the rate constants.}
   {Our detailed theoretical study of the CH + S $\rightarrow$ CS + H reaction shows that its rate is nearly independent
     on the temperature in a range of 10$-$ 500 K with an almost constant 
     value of   5.5 $\times$ $10^{-11}$ cm$^3$ s$^{-1}$ at temperatures above 100~K. This is a factor $\sim$ 2$-$3 lower
     than the value obtained with the capture model. 
     The rate of  the reaction C$_2$ + S $\rightarrow$ CS + C does depend on the temperature
     taking values close to 2.0 $\times$ $10^{-10}$ cm$^3$  s$^{-1}$ at low temperatures and increasing to 
 $\sim$ 5.0 $\times$ $10^{-10}$  cm$^3$  s$^{-1}$ for temperatures higher than 200~K.
 In this case, our detailed modeling taking into account the electronic and spin states provides a rate higher
than the one currently used by factor of $\sim$2. 
}
{  These reactions were selected for involving open-shell species with many degenerate electronic states,
  and, unexpectedly, the results obtained  in the present detailed calculations provide values which differ a factor
  of $\sim$2$-$3 from the simpler classical capture method.  We have updated the sulphur network with these new rates and compare
  our results in the prototypical case of TMC1 (CP). We find a reasonable agreement between model predictions
  and observations with a sulphur depletion factor of 20 relative to the sulphur cosmic abundance. However,
  it is not possible to fit all sulphur-bearing molecules better than a factor of 10 at the same chemical time.
 }
  

  \maketitle


  \section{Introduction} \label{sec:introduction}
  Astrochemistry has become a necessary tool for understanding the interstellar medium of our Galaxy and external galaxies.
  Nowadays, we are aware of the existence of nearly 300 molecules in the interstellar and circumstellar medium,
  as well as of around 70 molecules in external galaxies (for a complete list,
  see the Cologne Database for Molecular Spectroscopy\footnote{https://cdms.astro.uni-koeln.de/}).
  Although the sulphur cosmic elemental abundance is only ten times lower than that of carbon (S/H$\approx$1.5 $\times$ 10$^{-5}$),
  only 33 out of the currently detected interstellar molecules contain sulphur atoms.
  This apparent lack of chemical diversity in astrophysical sulphur-bearing
  molecules is the consequence of a greater problem in astrochemistry: there is an unexpected paucity
  of sulphur-bearing species in dense molecular clouds and star-forming regions.
  In such dense regions, the sum of the observed gas-phase abundances of sulphur-bearing species
  (the most abundant are SO, SO$_2$, H$_2$S, CS, HCS$^+$, H$_2$CS, C$_2$S, C$_3$S, and {NS} )
  constitutes only $<$1\% of the expected amount
  { \citep{Agundez2013, Vastel2018, Riviere2019, Hily-Blant-etal:22} }.
  One could think that most of the sulphur is locked on the icy grain mantles,
  but a similar trend is encountered within the solid phase, where s-OCS ($"$s-$"$ indicates that the molecule is in the solid phase) and s-SO$_2$ are the
  only sulphur-bearing species detected thus far \citep{Palumbo1995, Palumbo1997, Boogert1997, Ferrante2008},
  and only upper limits to the  s-H$_2$S abundance have been derived \citep{Jimenez2011}. { Recent
  observations with James Webb Space Telescope (JWST) did not detect s-H$_2$S either \citep{McClure2023}}.
  According to these data, the abundances of the observed icy species account for $<$ 5\% of
  the total expected sulphur abundance.
  This means that 94\% of the sulphur is missing in our counting.
  It has been suggested that this missing sulphur may be locked in hitherto undetected reservoirs
  in gas and icy grain mantles, or as refractory material \cite{Shingledecker2020}.
  In particular, laboratory experiments and
  theoretical work shows that sulphur allotropes, such as S$_8$, could be an important
  refractory  reservoir \citep{Jimenez2014,Shingledecker2020,Cazaux2022}.

  Gas phase Elemental abundances in Molecular CloudS (GEMS) is an IRAM 30m Large Program designed
  to estimate the S, C, N, and O depletions and the gas ionization fraction as a function of visual extinction
  in a selected set of prototypical star-forming filaments in low-mass (Taurus), intermediate-mass (Perseus),
  and high-mass (Orion) star forming regions \citep{Fuente2019,Navarro2020,Bulut-etal:21,Rodriguez-Baras2021,Esplugues2022,Spezzano2022,Fuente2023}.
  Determining sulphur depletion is probably
  the most challenging goal of this project.  The direct observation of the potential main sulphur reservoirs
  (s-H$_2$S, s-OCS, gas-phase atomic S) remains difficult even in the JWST era.
  Therefore, sulphur elemental abundance needs to be estimated by comparing the observed abundances
  of rarer species such as CS, SO, HCS$^+$, H$_2$S, SO$_2$, and  H$_2$CS,
   with the predictions of complex gas-grain chemical models. All this causes that
  the sulphur chemistry in cold dark clouds remains as a puzzling problem.
  The development of accurate and complete sulphur chemical networks is therefore a requisite
  to disentangle the sulphur elemental abundance. Within the context of GEMS project,
  we have carried out a large theoretical effort to improve the accuracy of key reaction rates
  of the sulphur chemical network, with a special interest in those associated with the formation
  and destruction paths of SO and CS which are the gas-phase sulphur
  bearing species observed in more different environments. We estimated the rates of the reactions
  S + O$_2$ $\rightarrow$ SO + O \citep{Fuente2016} and SO + OH $\rightarrow$ SO$_2$ + H  \citep{Fuente2019}
  at the low temperatures prevailing in dark clouds.  These reactions drives the SO chemistry in these cold environments.
  \citet{Bulut-etal:21} estimated the rate constant of  the CS + O  $\rightarrow$ SO + O reaction,
  that had been proposed as efficient CS destruction mechanisms in molecular clouds. 
  In this work we shall study two formation reactions of CS, which are thought to be important in regions with 
  low ionization fraction: CH($^2\Pi$)+S($^3$P) and C$_2$($^1\Sigma_g^+$)+ S($^3$P).
  In the two cases, the two reactants are radicals presenting several degenerate or quasi-degenerate
  electronic states: for CH + S there are 36 degenerate states and for C$_2$+S the first excited
  C$_2$($^3\Pi_u$) states are only 0.089 eV above the  C$_2$($^1\Sigma_g^+$) ground state . This
  makes  the experimental determination of their rates difficult, because of the low
  densities in which two radical species are obtained, and because of the possibility of self-reactions.
  Thus,  the most accurate theoretical determination of the reaction rate constants
  is desirable to proper bound  the abundance of CS in chemical models.

  The reactions rates currently available for these two reactions were obtained with a 
  classical capture method \citep{Vidal-etal:17}.  
  Dealing with open-shell systems, there are several degenerate electronic states for the reactants,
  not all of them leading to the desired product, and the crossing among them  originate barriers. All these effects
  are ignored in the classical capture method.  Precise calculations like the ones in this article are needed to improve the
  accuracy of our chemical networks and to estimate the uncertainties associated with the less demanding classical capture methods.

  \section{CH+S reaction}\label{sec:ch+s}
  
  The reaction
  \begin{eqnarray}\label{Eq:CH+S}
    \textrm{CH}(^2\Pi)+\textrm{S}(^3\textrm{P}) &\rightarrow& \textrm{CS}(X^1\Sigma^+,a^3\Pi)  + \textrm{H} \quad\quad (a)\nonumber \\
    \\
    &\rightarrow& \textrm{SH}(X^2\Pi)  + \textrm{C}(^3\textrm{P}) \quad\quad (b)\nonumber
  \end{eqnarray}
  presents several rearrangement channels (CS and SH products) with several electronic states
  in each case. There are 36 degenerate states (neglecting spin-orbit couplings) in the
    CH($^2\Pi$)+S($^3P$) entrance channel. The same number of degenerate states are
   in the SH($X^2\Pi$)  + C$(^3P)$ rearrangement channel, which is only $\approx$ 0.08 eV below
  CH($^2\Pi$)+S($^3P$). The reaction SH($X^2\Pi$)  + C($^3P$)$\rightarrow$
  CS(X$^1\Sigma^+,a^3\Pi$)  + H has already been studied theoretically for the ground $^2A'$ state
  \citep{Stoecklin-etal:88,Stoecklin-etal:90a,Stoecklin-etal:90b,Song-etal:16,Voronin:02}
  and for the excited $^2A''$ state \citep{Zhang-etal:18b}.

  The 36 degenerate electronic states (neglecting spin-orbit couplings) involved in Eq.~(\ref{Eq:CH+S}.a)
  consist of 6 spin states times 6 orbital states.
  The spin states are quartet and doublets, while the orbital states are splitted in A' and A'' states, $i.e.$,
  symmetric or antisymmetric with respect to the inversion through the plane of the molecule.
  Of all these 36 states, only a doublet (i.e. 2 states) correlates to the CS($X^1\Sigma^+$)
  states, which correspond to the ground adiabatic $^2A'$ states. Reaction ~(\ref{Eq:CH+S}.a)
towards the $X^1\Sigma^+$ is exothermic by $\approx$ 3.5 eV. However, the excited CS($a^3\Pi$)
is about 3.42 eV above  CS($X^1\Sigma^+$), and the reaction is nearly  thermoneutral.

  \subsection{Ab initio calculations}

  To describe the electronic correlation along the reaction, we use
  here the internally contracted multi-reference configuration interaction (ic-MRCI)  method
 \citep{Werner-Knowles:88,Werner-Knowles:88b} including the Davidson correction (hereafter called MRCI+Q) \citep{Davidson:75}
 and the calculations are performed with the MOLPRO suite of programs \citep{MOLPRO-WIREs}. Three electronic states are calculated
 for each symmetry, 3  $^2A'$ and 3 $^2A''$, and the same for the quartet states.

 In these calculations, the molecular orbitals are optimized using
 a state-averaged complete active space self-consistent field (SA-CASSCF) method,
 with an active space of 10 orbitals (7 and 3  of  $a'$ and $a''$ symmetry, respectively).
 Five $^{2,4}A'$ and four $^{2,4}A''$ electronic states are calculated 
and simultaneously optimized at CASSCF level, using a dynamical weighting
 factor of 10. In all these calculations the aug-cc-pVTZ (aVTZ) basis set is used \citep{Dunning:89}.
For the ic-MRCI calculations, 6 orbitals are kept doubly occupied,
giving rise to $\approx$ $5 \times 10^6$ $(382 \times 10^6)$
contracted (uncontracted) configurations. Checks made with the aug-cc-pVQZ basis  gave nearly parallel results along some
of the minimum energy paths (MEPs) shown below. For this reason, we kept the aVTZ basis set
to build the PES. Nevertheless, to better describe the long range part, we shall use a AV5Z basis, as discussed below.

\begin{figure}[t]
\begin{center}
  \includegraphics[width=7.cm]{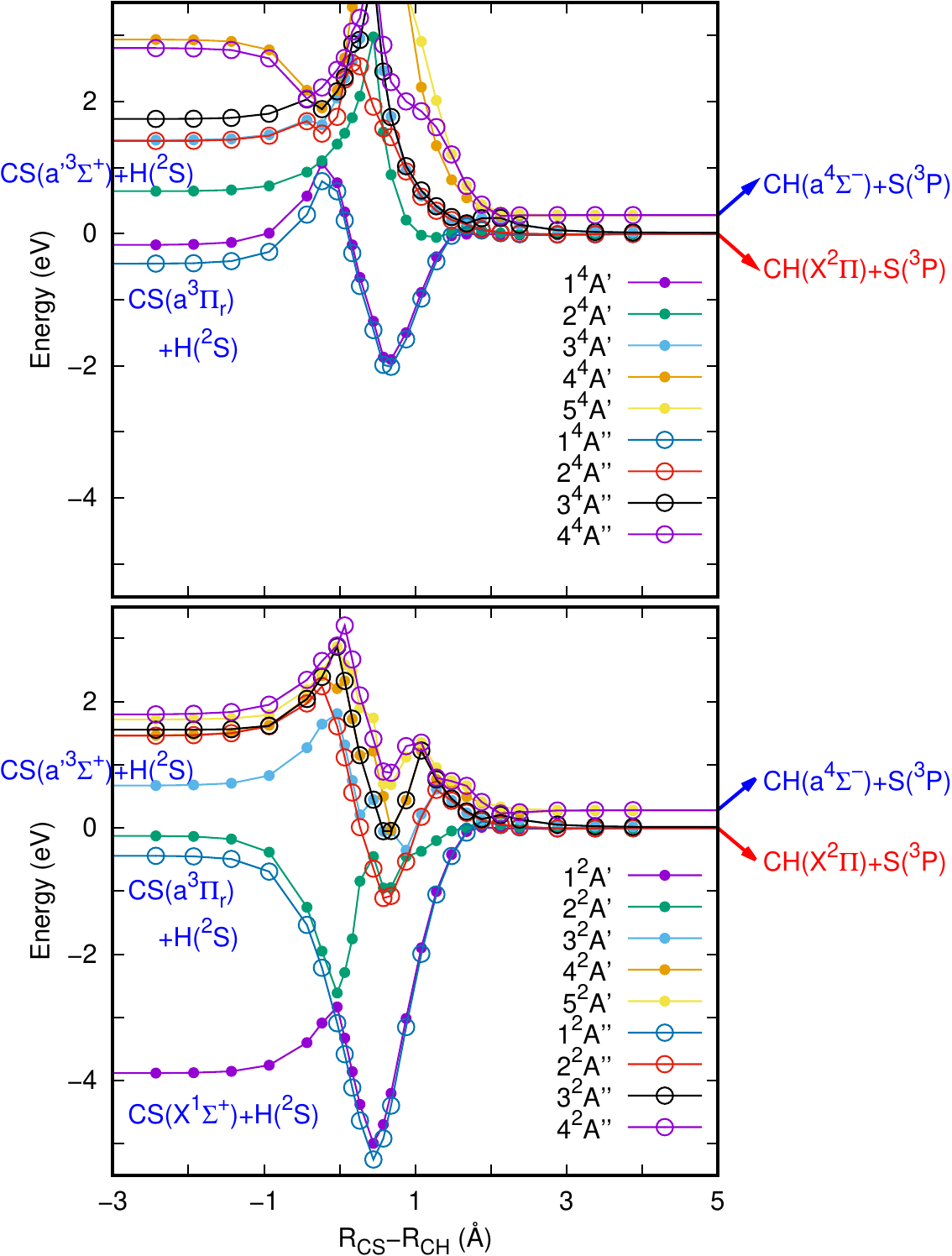}
  
  \caption{\label{fig:mep-cas}{ CASSCF energies along the reaction coordinate defined as R$_{CS}$-R$_{CH}$ distance
      difference, obtained for an angle $\gamma_{SCH}$= 179.9$^o$ for the doublet (bottom panel) and quartet (top panel) states,
       for  the ${\rm CH}+ {\rm S}  \rightarrow {\rm CS}+ {\rm H}$ reaction. Five A' and 4 A'' electronic states are considered
  }}
\end{center}
\end{figure}

In Fig.~\ref{fig:mep-cas} the first 5 A' and 4 A'' states for doublet (bottom) and quartet (top) multiplicities
are shown, calculated at CASSCF level.
With no spin-orbit couplings, the six electronic states for each multiplicity are degenerate for long distances
between S($^3P$) and CH($X^2\Pi$), and this asymptote is taken as the origin of energy.
When they approach, they cross with the excited states correlating to
S($^3P$)+CH($a^4\Sigma^-$). After the crossing, there are three (two) curves for the doublet (quartet) states
that become negative. For the doublets, one of these states
correlates to  CS($X^1\Sigma^+$) and two, nearly degenerate, to CS($a^3\Pi_r$) states.

\begin{figure}[t]
\begin{center}
 \includegraphics[width=7.cm]{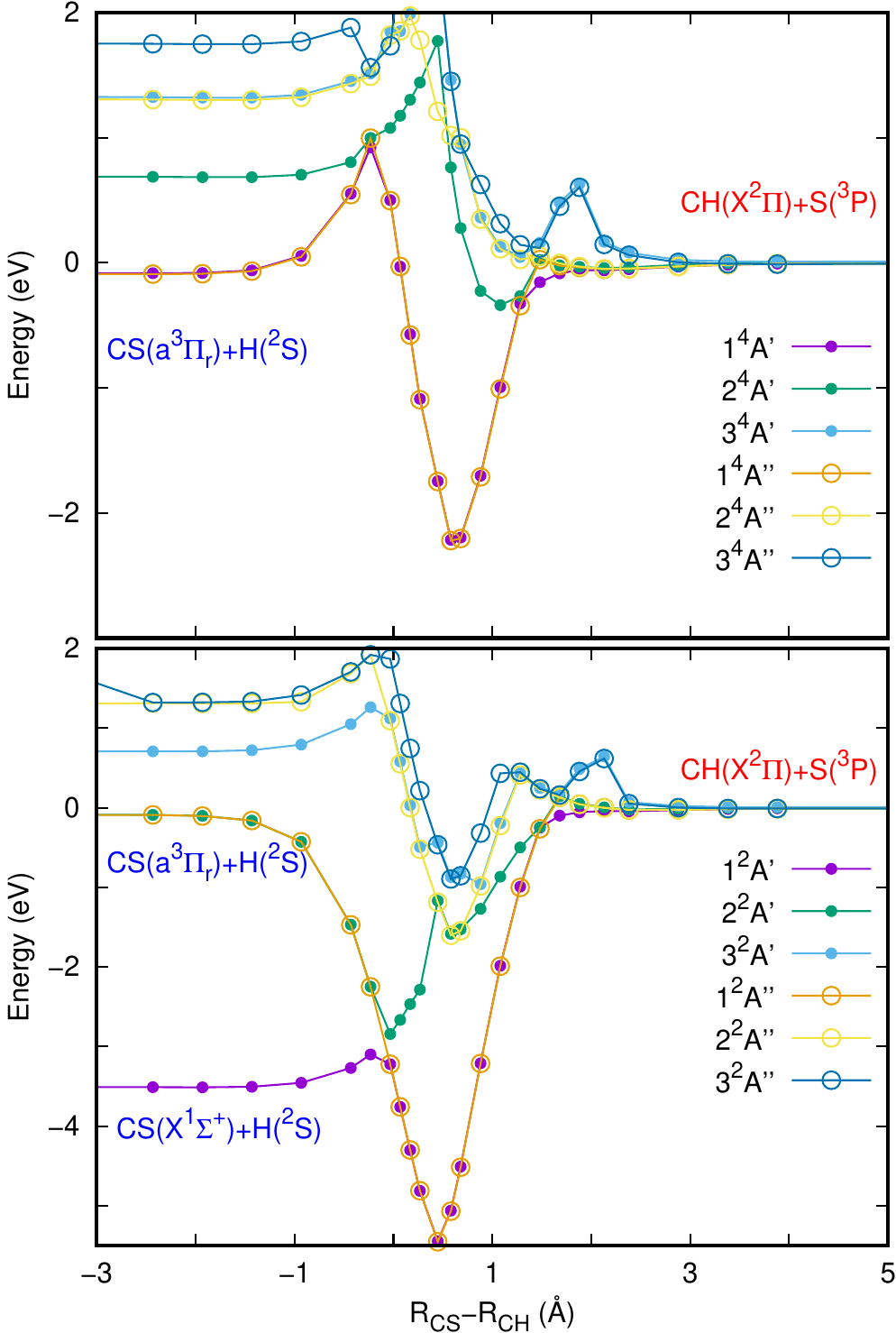}  
  \caption{\label{fig:mep-mrci}{ MRCI+Q energies along the reaction coordinate defined as R$_{CS}$-R$_{CH}$ distance
      difference, obtained for an angle $\gamma_{SCH}$= 179.9$^o$ for the doublet (bottom panel) and quartet (top panel) states,
       for  the ${\rm CH}+{\rm S}  \rightarrow {\rm CS}+ {\rm H}$ reaction. Three A' and three A'' electronic states are considered in each case.
  }}
\end{center}
\end{figure}

This degeneracy is recovered in the  MRCI+Q calculations shown in Fig.~\ref{fig:mep-mrci}. The two
degenerate states, 1$^2A'$ and 1$^2A''$ (and  1$^4A'$ and 1$^4A''$ for quartets),
correspond to the HCS($^2\Pi$) radical (and the excited HCS($^4\Pi$)) state, which experience
Renner-Teller effects, as studied by \cite{Senekowitsch-etal:90}. In the case of doublets,
the $X^2A'$ state (of $^2\Pi$ character at collinear geometry) crosses with  $2^2A'$ (of $^2\Sigma$
character at collinear geometry), which correlate with the CS(X$^1\Sigma^+$) state of the products.

The crossing at collinear geometry is a conical intersection, between 
$\Sigma$ and $\Pi$ states. As long as the system bents, there are couplings and the crossing
is avoided. As a consequence, the ground $X^2A'$ state correlates to the CS(X$^1\Sigma^+$) products,
$i.e.$ only these 2 doublets among the  36 states correlating to the CH($^2\Pi$)+S($^3P$) reactants.
Therefore, only the ground electronic state is needed to describe
the reaction \ref{Eq:CH+S}.
{The energy difference between the CH($^2\Pi$)+S($^3P$)
  and CS(X$^1\Sigma^+$) + H, at their corresponding equilibrium geometries is $D_e$ = 3.52 eV,
  and $D_0$ = 3.62 eV  when including zero-point energy (using the fit described below).
  Since there is no direct measurements for this reaction, the experimental exothermicity
  is estimated from the dissociation energies
D$_0$=D$^{CH}_0$ - D$^{CS}_0$ = 3.89 eV, about a 10\% higher than the present results
using the values reported by \cite{Herzberg-etal:79}.
}

\subsection{Long range interaction}

MRCI calculations are not size consistent and the Davidson correction (+Q) is not adequate
to describe close lying electronic states. Therefore MRCI+Q method introduces inaccuracies
in the long range region, which need to be described accurately to obtain good rate constants
at low temperatures.
Coupled Cluster methods are size consistent and are typically considered as a holy grail to describe long range interactions,
but in the presence of two open shell reactants, is not expected
to yield good accuracy either.

\begin{figure}[t]
\begin{center}
  \includegraphics[width=7.cm]{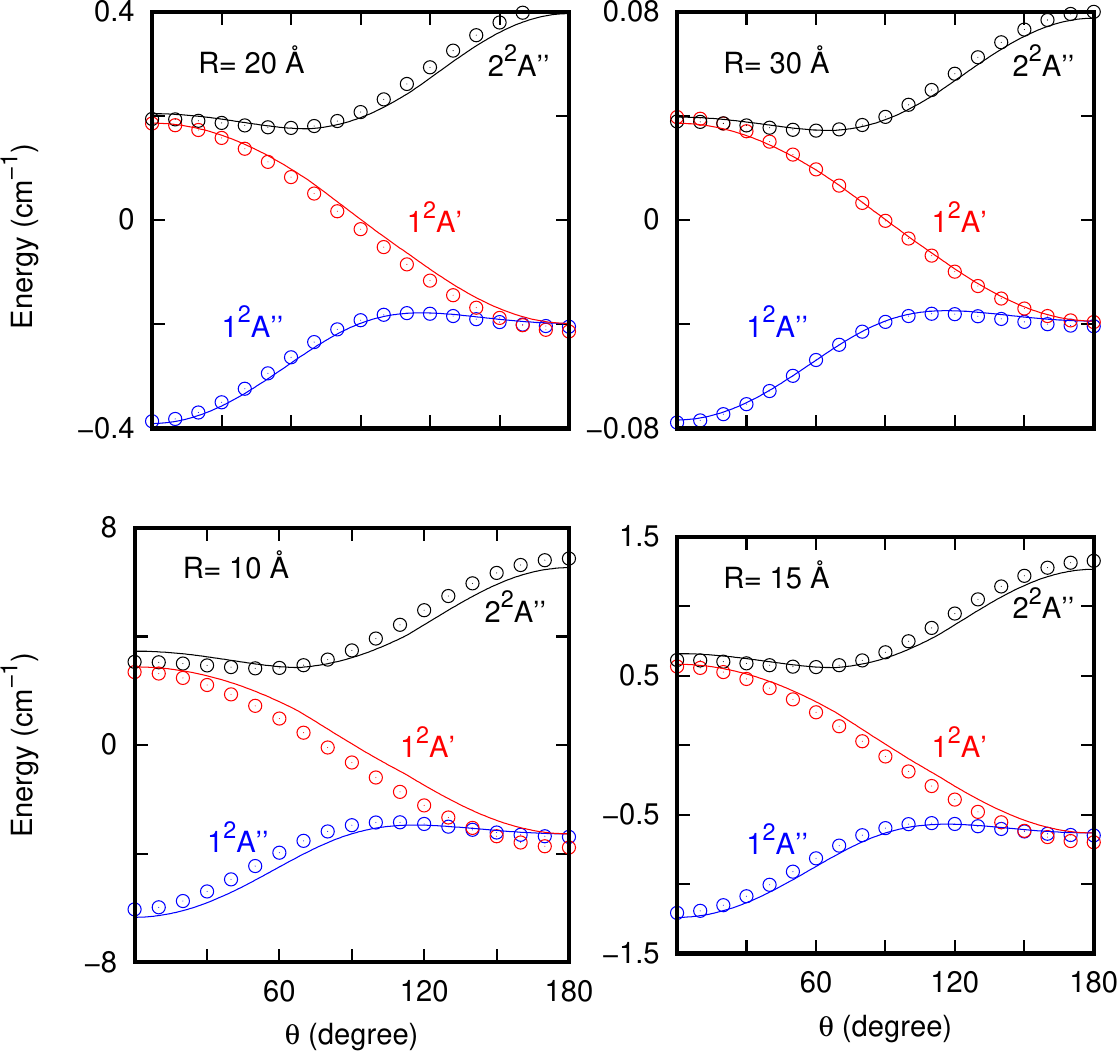}
  
  \caption{\label{fig:long-range}{ Points are the CASSCF(aV5Z) energies (in cm$^{-1}$) as a function of the Jacobi angle $\theta$ (defined
      in the text) for $r_{CH}$= 1.1199\AA\, and $R$= 10, 15, 20 and 30\AA, as indicated in each panel, for the $1^2A''$, $1^2A'$ and $2^2A''$ states.
      Lines correspond to the long range analytical fit of Eq.~(\ref{Eq:long-range}), taken from \citep{Zeimen-etal:03}.
  }}
\end{center}
\end{figure}

As an alternative, here we use the CASSCF method with a larger  aV5Z basis set \citep{Dunning:89}.
Several points have been calculated in Jacobi coordinates,  for CH($r_{CH})$ =  1.1199 \AA\ and R= 10-50 \AA\, where $R$ is the distance
of CH center-of-mass and the S atom, as a function of the angle $\theta$, with $\cos\theta={\bf R}\cdot {\bf r}_{CH}/ R r_{CH}$.
 At long distances, the system behaves as dipole-quadrupole interaction, with the dipole of CH 
interacting with the quadrupole of a P sulphur atom, whose analytical form is \citep{Zeimen-etal:03}
\begin{eqnarray}\label{Eq:long-range}
  V_{LR}(R,\theta)= M_{QD}(\theta) {Q_A d_B\over R^4} + M_{QQ}(\theta) {Q_A Q_B\over R^5},
\end{eqnarray}
with $M_{QD}(\theta)$ and $M_{QQ}(\theta)$ being 3$\times$3 matrices, depending on Legendre
polynomials \citep{Zeimen-etal:03}. The eigenvalues of this matrix properly describe 
the angular dependence of the  adiabatic states. Thus, only 2 effective parameters
are needed to fit the {\it ab initio} points, $Q_A d_B$ = 0.293 hartree \AA$^{-4}$ and $Q_A Q_B$ = 0.092 hartree \AA$^{-5}$.
Two families of three states are considered, separately, $1^2A'', 1^2A', 2^2A''$ and $2^2A', 3 ^2A',3 ^2A''$, 
corresponding to the two $\Pi$ states of CH interacting with the 3 $P$ states of sulphur.
The excellent agreement between calculated and fitted analytical expressions, as shown
in Fig.~\ref{fig:long-range}, demonstrates the adequacy of the analytical fit in the asymptotic region.

\subsection{Analytical fit of the ground state}

The analytical representation of the potential energy surface of the ground 1$^2A'$ electronic
state is described by two terms
\begin{eqnarray}\label{pes-partition}
  V(r_{CH},r_{CS},r_{SH})= E_g^{FF}(r_{CH},r_{CS},r_{SH}) + V^{3B}(r_{CH},r_{CS},r_{SH}),
\end{eqnarray}
where $V^{3B}$ is the three-body term added to the zero order description provided by
the lowest root, $ E_g^{FF}(r_{CH},r_{CS},r_{SH})$, of the 3$\times$3 reactive force-field matrix
defined as \citep{Zanchet-etal:18,Roncero-etal:18,Goicoechea-etal:21}
\begin{eqnarray}
  H^{FF} =\left({\scriptsize
  \begin{array}{ccc}
      V_{CH} + W^1_{CS} + W^1_{SH}+V_{LR} &  V_{12}   &  V_{13}\\ 
      V_{12} & V_{CS} + W^2_{CH} + W^2_{SH} &  V_{23}   \\ 
      V_{13} & V_{23} &  V_{SH} + W^3_{CH} + W^3_{CS}   
  \end{array}
 }\right).
\end{eqnarray}
The diagonal terms describe each of the rearrangement channels, in which 
  $V_{CH}(r_{CH})$, $V_{CS}(r_{CS})$ and $V_{SH}(r_{SH})$, are fitted using the diatomic terms
  of \cite{Aguado-Paniagua:92}.
  In the reactants channel 1, the long-range term given by Eq.~\ref{Eq:long-range} is included. 
  $W_{AB}$ are Morse potentials whose parameters are determined to describe each channel independently.
  Finally, the non-diagonal terms $V_{ij}$ essentially are build as Gaussian functions $exp(-\alpha (H^{FF}_{ii} -H^{FF}_{jj})^2)$
  depending on the energy difference between the diagonal terms of the corresponding force-field matrix. The parameter $\alpha$
  is determined to fit the transition between rearrangements.

  The three-body term, $V^{3B}$ in Eq.~(\ref{pes-partition}), is described with the method of \cite{Aguado-Paniagua:92}
  using a modification of the program GFIT3C \citep{Aguado-etal:98}. In the fit, about 7500 {\it ab initio} points,
  calculated at MRCI+Q level, have been included. These points are mainly composed by a grid of 22 points
  for $0.6 \leq r_{CH}\leq $ 8 \AA, 27 points for 1$\leq r_{CS}\leq$ 8 \AA, and 19 points in the $\theta_{HCS}$ angle,
  in interval of 10$^o$. The points have been weighted, with a weight of 1, for those with energy up to 1 eV, from the entrance
  CH+S channel at 50 \AA, and with a Gaussian function for higher energies, and considering a minimum weight of 10$^{-4}$.
  The final fit uses polynomials up to order 10, with an overall root-mean-square error of 0.07 eV.

\begin{figure}[h]
\begin{center}
  \includegraphics[width=8.cm]{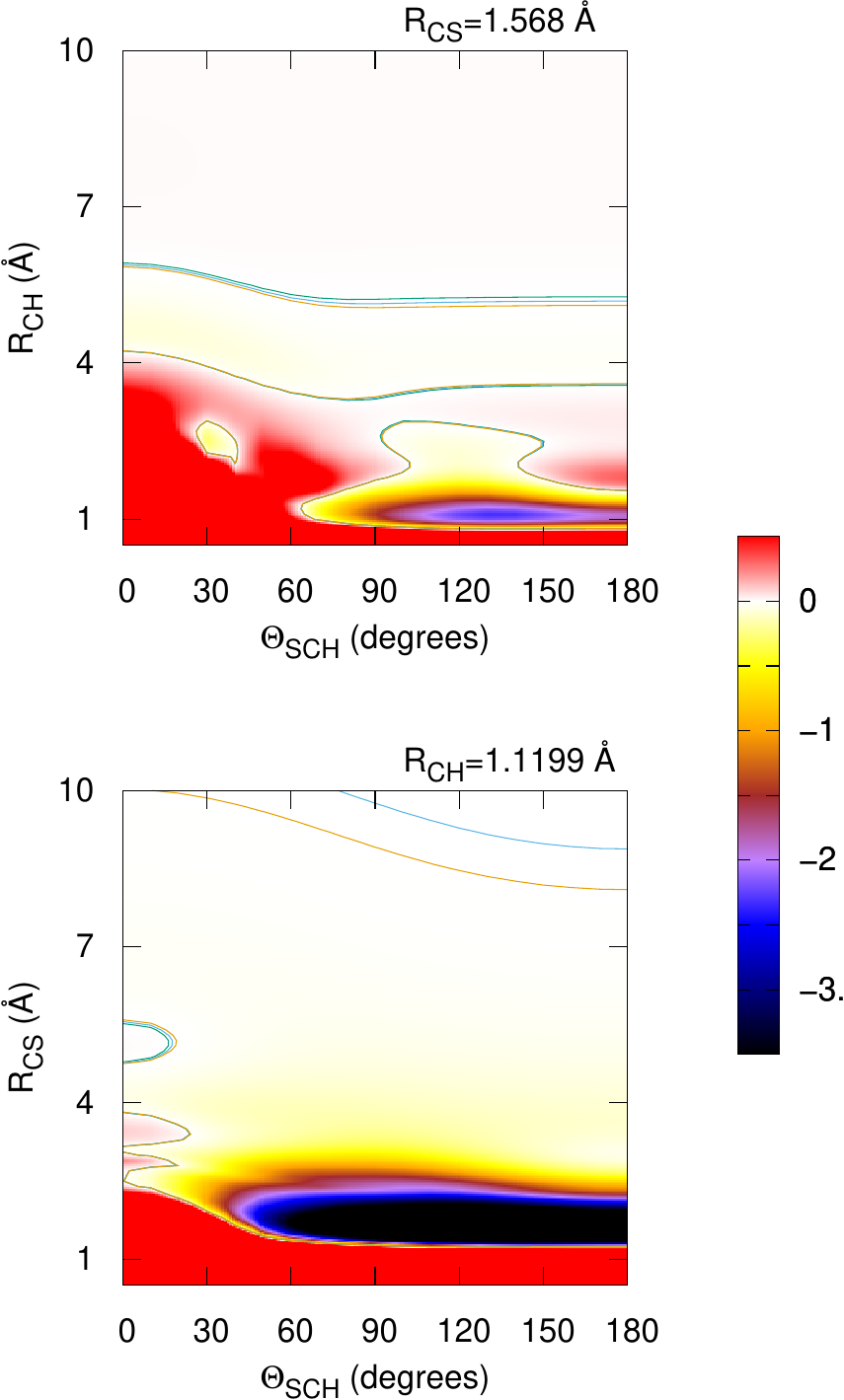}
  
  \caption{\label{fig:pes}{Contour plots of the analytical potential energy surfaces fitted to {\it ab initio} points,
      corresponding to the reactant channel for  $R_{CH}$=1.1199 \AA (bottom panel) and to the products channel
      for $R_{CS}$=1.568 \AA, . Energies are in eV, and for the products the energies are shifted  3.526 eV, so that
      its zero of energy corresponds to the CS($r_{eq}$) at an infinite distance from H. The contours correspond
      to 0 and $\pm$ 1 meV, to show the dependence of the potential at long distances. The units of the color box is in eV.
  }}
\end{center}
\end{figure}

The main features of the potential fit is shown in Fig.~\ref{fig:pes}, where the contour plots
for the reactants (bottom panel, for $R_{CH}$=1.1199 \AA) and products (top panel, for $R_{CS}$=1.568 \AA, and shifted by 3.526 eV,
the exoergicity of the reaction) are shown.
The CH+S reactant channel (bottom panel) is attractive for $R_{CS} < $ 7 \AA, leading to
the products channel at $R_{CS}\approx$ 1.5 \AA, with an energy of  -5.5 eV. These energies correspond to the
CS-H well in products channel, which is about 2 eV below the CS + H products, shown in the top panel of Fig.~\ref{fig:pes},
shifted 3.526 eV to show the details. At $R_{CH}\approx$ 2\AA, there is a barrier which arises from the curve crossing
discussed above.

\subsection{Quasi-Classical versus Quantum wave packet  dynamics in the $1^2A'$ state}\label{sec:dyn}

To check the validity of the quasi-classical method, we first compare quantum and quasi-classical trajectory (QCT) calculations
for total angular momentum $J$=0. 
The quantum wave packet (WP) calculations are performed with the MADWAVE3 code \citep{Zanchet-etal:09b} and the parameters
used are listed in Table ~\ref{wvp-parameters}. The WP method is considered numerically exact, but
it is very demanding computationally. The QCT calculations are performed with the MDwQT code \citep{Sanz-Sanz-etal:15,Zanchet-etal:16,Ocana-etal:17}.
 Initial conditions are sampled with the usual
 Monte Carlo method \citep{Karplus-etal:65}. In this first set of calculations CH is in its ground  vibrational (v) and rotational
 state, (v,j)=(0,0), and  the initial internuclear distance and velocity distributions are
 obtained with the adiabatic switching method \citep{Grozdanov-Solovev:82,Qu-Bowman:16,Nagy-Lendvay:17}. An impact parameter
 $b=0$ is considered, that corresponds to $J$=0. The initial distance between sulfur and the CH center of mass
 is set to 50 Bohr, and the trajectories are stopped when any internuclear distance is longer than 60 Bohr.
 For each energy $N_{tot}=$ 10$^4$ trajectories are run to calculate the reaction probability
 as $P_R(E)= N_r /N_{tot}$, where $N_r$ are the number of reactive trajectories. 
 
\begin{figure}[t]
\begin{center}
 \includegraphics[width=8.cm]{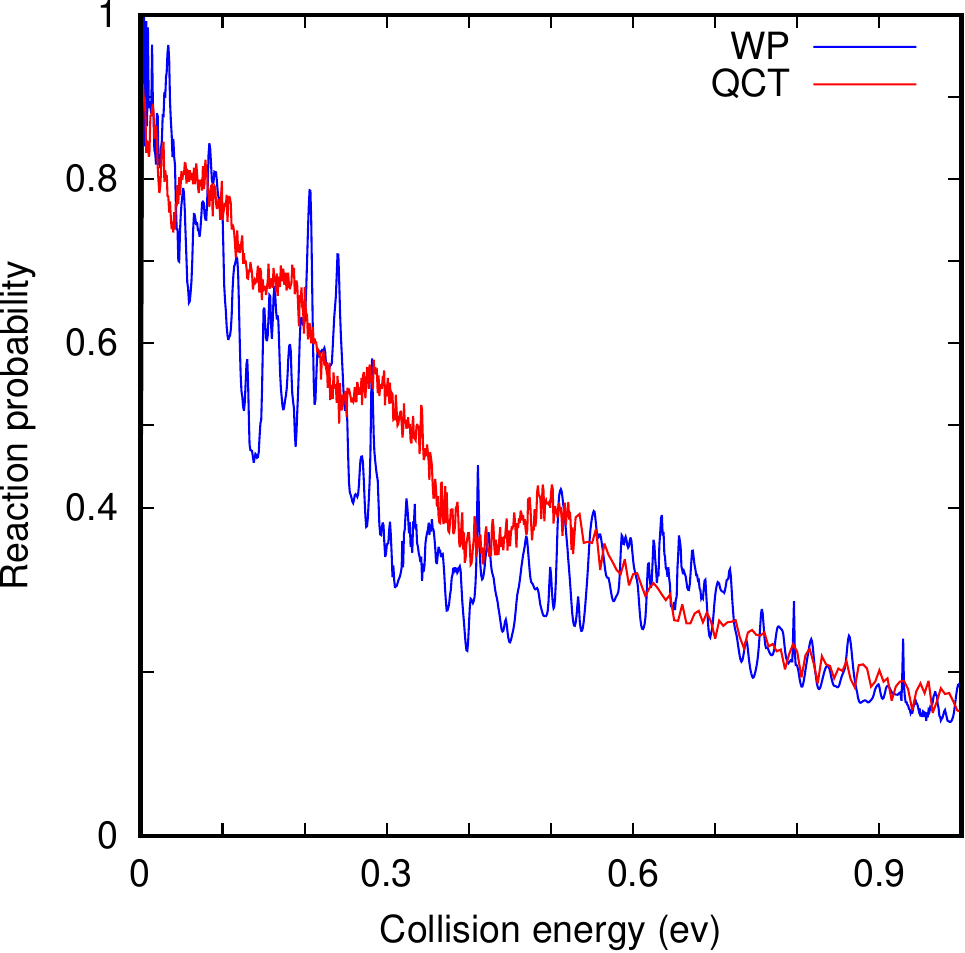}
  
 \caption{\label{probJ0}{CH+S $\rightarrow$ CS + H reaction probabilities versus collision energy 
     for $J=0$ in the 1 $^2A'$ electronic state using  quantum wave packet (WP)  and
     quasi-classical trajectory (QCT)
     methods.
  }}
\end{center}
\end{figure}

 The WP and QCT reactions probabilities are compared in Fig.~\ref{probJ0}, and the two show
 methods show a very similar behavior:  a reaction probability slightly larger than 0.9
 at energies below 0.01 eV, decreasing as a function of collision energy, down to a probability lower
 than 0.2 at 1 eV. In the two cases there are oscillations, which do not match perfectly but
 that show similar envelopes. Since these oscillations are expected to wash out when considering
 the partial wave summation over total angular momentum, $J$, we consider this agreement as satisfactory.
 This leads us to conclude, that QCT method is sufficiently accurate to determine the reaction rate constant,
 as described below.

%
 \begin{table}[h]
 \caption{\label{wvp-parameters}
   Parameters used in the wave packet calculations in reactant Jacobi coordinates:
   $r_{min} \leq r\leq r_{max}$ is the CH internuclear distance, $R_{min} \leq R\leq R_{max}$ is the distance between CH
   center-of-mass and the sulphur
   atom, $0 \leq \gamma \leq \pi$ is the angle between ${\vec r}$ and ${\vec R}$ vectors. The initial wave packet
   is described in $R$ by a Gaussian centered at $R=R_0$, and at a translational energy of
   $E=E_0$, and width $\Delta E $. The total reaction probability is obtained by analyzing the total flux
   at $r=r_\infty$.
}
 \begin{center}
 \begin{tabular}{|cc|}
 \hline 
 $r_{min}$, $r_{max}=$  0.1, 17 \AA & $N_r$=420 \\
 $r_{abs}$=  5 \AA & \\
$R_{min}$, $R_{max}=$   0.001, 17.5\AA & $N_R$=840  \\
 $R_{abs}$=  10.5 \AA  &\\
$N_\gamma$ = 168 & in $[0,\pi]$  \\
$R_0$  = 10 \AA & $E_0,\Delta E$= 0.35,0.2 eV\\
$r_\infty$ = 4 \AA &   \\
 \hline
 \end{tabular}
 \end{center}
 \end{table}

 The reaction rate constant for CH($v$=0 and 1) in the 1$^2A'$ electronic state
 is evaluated according to  \begin{equation}
  K_v^{1^2A'}(T) =  \sqrt{\frac{8 k_B T}{\pi \mu}} \pi b_{max}^2(T)P_r(T)
  \label{eqn:k_T},
\end{equation}
where $b_{max}(T)$ and $P_r(T)$ are the maximum impact parameter and
reaction probability at constant temperature, respectively. In this case, about 10$^5$ trajectories
are run for each temperature, the same for translation and rotation degrees of freedom,
fixing the vibrational state of CH to $v$=0 or 1.

 \subsection{Thermal rate}

\begin{figure}[t]
\begin{center}
  \includegraphics[width=7.cm]{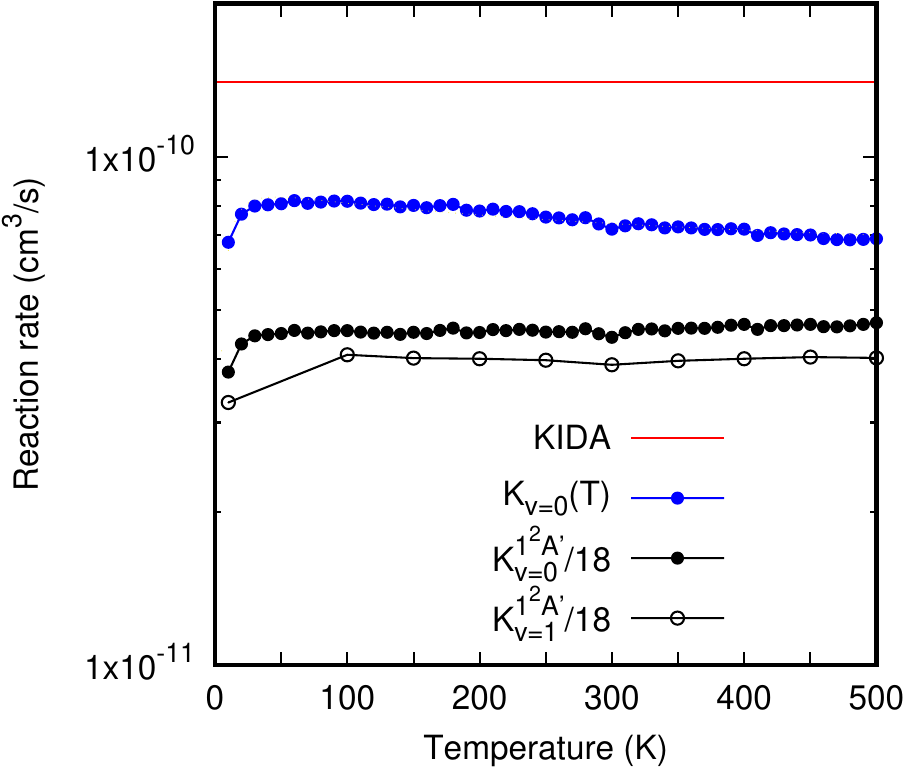}
  
  \caption{\label{fig:CH+S-thermal-rate}{Vibrational-selected rate constant for the CH(X$^2\Pi$,v=0)+ S($^3P$) $\rightarrow$ CS($X^1\Sigma$) + H reaction,
      obtained here according to Eq.~(\ref{Eq:CH+S-thermal-rate}).  
  }}
\end{center}
\end{figure}

 Considering  that only the double degenerate 1$^2A'$ electronic state reacts to form CS($X^1\Sigma^+$),
 the electronic partition function has to be considered. Neglecting  the spin-orbit splitting,
 the electronic partition function would be 2/36, $i.e.$, the thermal rate constant is about 1/18 the
 reaction rate, $K^{1^2A'}$, associated to the 1$^2A'$ state.
 Including the spin-orbit splitting of S($^3P_{J_S}$) 
 sulphur atom, and assuming that only the lowest two spin-orbit react (having
 an individual rate constant equal to $K^{1^2A'}$), the vibrational selected thermal rate constant is given by
 \begin{eqnarray}\label{Eq:CH+S-thermal-rate}
   K_v(T) =  {2 K_v^{1^2A'}(T) \over 4\left[ 5 + 3 exp(-539.83/T) + exp(-825.34/T)\right]},
 \end{eqnarray}
 and it is shown in Fig.~\ref{fig:CH+S-thermal-rate}.
In this figure the red line corresponds to the rate constant (1.4 $\times$ $10^{-10}$ cm$^3$s$^{-1}$)
obtained from the KIDA data base, as obtained with a classical capture model \citep{Vidal-etal:17},
using analytical formulas \citep{Georgievskii-Klippenstein:05,Woon-Herbst:09} with dipole
moments and polarizabilities taken from the literature or calculated using Density Functional theory.

The  $K_{v=0}^{1^2A'}(T)/18$ rate constant  at 10 K is about
4  $\times$ $10^{-11}$ cm$^3$/s, increasing to a nearly constant value of
$\approx$  5.5  $\times$ $10^{-11}$ cm$^3$/s at temperatures
above 100 K. This is a factor between 2 and 3 lower
than the value obtained with the capture model \citep{Vidal-etal:17}.

When including the spin-orbit splitting,
the rate $ K_{v=0}(T)$ is larger at 10 K, simply because the populations
of the excited sulphur spin-orbit states, $J_S$= 1 and 0, are negligible. As temperature increases, their populations
increase, leading to a reduction of the rate constant $K_{v=0}(T)$, which decreases tending to $K^{1^2A'}(T)/18$
at high temperature. The rate with spin-orbit splittings, $K_{v=0}(T)$, is only a factor $\approx$ 1/2 smaller than 
that of KIDA at 10 K, and is considered the most accurate obtained here. The $K_{v=0}(T)$ rate constant
has been fitted and the parameters are listed in  Table~\ref{tab:new-rates}.

In Fig.~\ref{fig:CH+S-thermal-rate}, $K_{v=1}^{1^2A'}(T)/18$ is also shown. Its contribution at low temperatures
is small, because the vibration energy of CH(v=1) is about 0.337 eV higher than v=0, $i.e.$ 3916 K.
Therefore, $K_{v=0}(T)$ is a good approximation to
the thermal rate constant, which includes the spin-orbit splitting.

 \section{C$_2$+S reaction}\label{sec:c2+s}

\begin{figure}
\centering
\includegraphics[angle=0,width=1\linewidth]{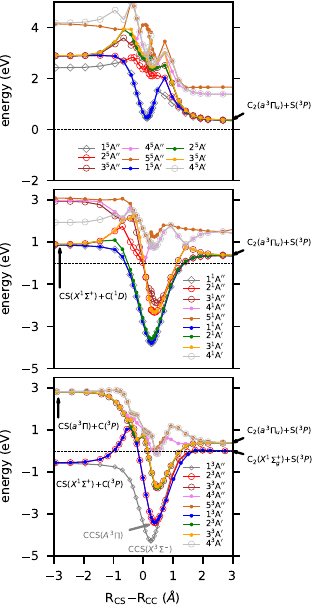}
\caption{\footnotesize CASSCF/aV$T$Z optimized reaction paths for 
the conversion between C$_2$+S and CS+C which occur on several potential energy surfaces of CCS. 
The reaction coordinate is defined as $R_{\mathrm{CS}}\!-\!R_{\mathrm{CC}}$, while the valence angle,
$\gamma_{\mathrm{CCS}}$, was held fixed at $179.9\si{\degree}$ in the geometry  optimizations. Five $A''$
and four $A'$ electronic states of CCS were considered for 
each multiplicity: triplet (bottom panel), singlet (middle panel) and quintet (top panel).}  
\label{fig:cas_c2}
\end{figure}

\begin{figure}
\centering
\includegraphics[angle=0,width=1\linewidth]{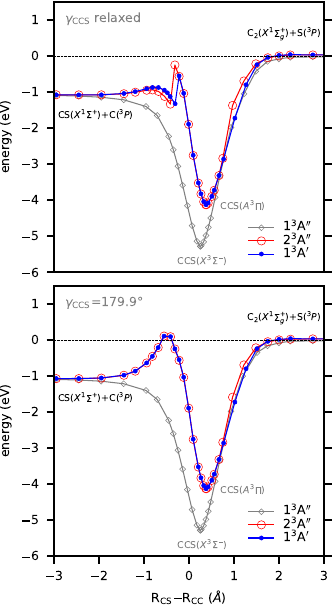}
\caption{\footnotesize Optimized reaction paths for
  $\mathrm{C_{2}}(X^{1}\Sigma_{g}^{+})\!+\!\mathrm{S}(^{3}P)\rightarrow\mathrm{CS}(X^{1}\Sigma^{+})\!+\!\mathrm{C}(^{3}P)$
  as obtained from MRCI+Q/aV$Q$Z//CASSCF/aV$Q$Z calculations.
  The reaction coordinate is defined as $R_{\mathrm{CS}}\!-\!R_{\mathrm{CC}}$.
  In the bottom panel, the valence angle, $\gamma_{\mathrm{CCS}}$,
  was held fixed at $179.9\si{\degree}$ in the geometry optimizations, while,
  in the top panel, it has been freely optimized along with $R_{\mathrm{CS}}$ or $R_{\mathrm{CC}}$
  bond distances. Two $^{3}A''$ and one $^{3}A'$ triplet electronic states of CCS were considered in each case.} 
\label{fig:mrci_c2}
\end{figure}

The reaction 
\begin{equation}\label{eq:c2_s}
\mathrm{C_{2}}(X^{1}\Sigma_{g}^{+})+\mathrm{S}(^{3}P)\rightarrow\mathrm{CS}(X^{1}\Sigma^{+})+\mathrm{C}(^{3}P) 
\end{equation}
is exothermic by $\sim\!1\,\si{\electronvolt}$~\citep{VIS019:1645,RED003:419}. 
Due to the open shell nature of the involved atoms [$\mathrm{S}(^{3}P)$~and~$\mathrm{C}(^{3}P)$],
reactants collision and product formation can take place adiabatically
on three triplet CCS potential energy surfaces ($2^{3}A''\!+\!1^{3}A'$); see,~\emph{e.g.},~Figures~\ref{fig:cas_c2}~and~\ref{fig:mrci_c2}. 
To the best of our knowledge, no
dedicated theoretical studies are yet available on this reaction. 
\citet{Vidal-etal:17} reported a theoretical upper limit for its rate coefficient
($k\!\sim\!\SI{2e-10}{}\,\rm cm^{3}\,molecule^{-1}\,s^{-1}$ at $10\,\si{\kelvin}$)
using classical capture rate theory. From an experimental viewpoint, presently
available techniques for the production of reactant dicarbon often generate a mixture
of both $\mathrm{C_{2}}(X^{1}\Sigma_{g}^{+})$ and $\mathrm{C_{2}}(a^{3}\Pi_{u})$~\citep{GU006:245,PAR008:9591}
(recall that this electronically excited state lies only $\sim\!716\,\mathrm{cm^{-1}}$
above the ground $X$ form) which, together with the expected high reactivity of C and S atoms,
make the laboratory characterization of this specific
reaction~(\ref{eq:c2_s}) extremely cumbersome.

\subsection{Ab initio calculations}\label{subsec:abinitio_c2}

The methodology employed to obtain optimized energy paths for the C$_{2}$+S$\rightarrow$CS+C reaction
closely resembles the one utilized for the CH+S system. Preliminary PES explorations and geometry optimizations
were all performed at the SA-CASSCF level of theory, followed by single-point MRCI+Q calculations. The CASSCF active space involves a total of 
14 correlated electrons in 12 active orbitals (9$a'$+3$a''$).
For each multiplicity considered (triplet, singlet and quintet), five $A''$ and four $A'$ electronic states
were simultaneously treated in the SA-CASSCF wave functions.
The aV$X$Z ($X\!=\!T,Q$) basis sets of Dunning and co-workers~\citep{DUN89:1007,KEN92:6796} were employed throughout,
with the calculations done with MOLPRO.  

Our calculated CASSCF/aV$T$Z optimized path for reaction~(\ref{eq:c2_s}) is shown
at the bottom panel of Figure~\ref{fig:cas_c2}.
As seen, the $\mathrm{C_{2}}(X^{1}\Sigma_{g}^{+})+\mathrm{S}(^{3}P)$ reactants
collision involves only triplet CCS PESs and can 
happen on two $^{3}A''$ and one $^{3}A'$ electronic states.
Proceeding through the ground-state PES of $\mathrm{CCS}(1^{3}A'')$,
reaction~(\ref{eq:c2_s}) does not encounter any  activation barriers for collinear atom-diatom approaches, being exothermic by $\sim\!0.6\,\si{\electronvolt}$ at CASSCF/aV$T$Z level. 
Note that this process occurs via the formation of a strongly-bound intermediate complex corresponding
to the linear global minimum of CCS, 
$\ell$-$\mathrm{CCS}(X^{3}\Sigma^{-})$~\citep{SAI87:L115}; from this structure, 
the ground-state $\mathrm{CS}(X^{1}\Sigma^{+})+\mathrm{C}(^{3}P)$ products can be directly accessed,
without an exit barrier. 

A close look at Figure~\ref{fig:cas_c2} (bottom panel) also reveals that the excited $2^{3}A''$ and $1^{3}A'$ electronic states are degenerate along $C_{\infty v}$ atom-diatom 
collisions. These PESs form the Renner-Teller components of the strongly-bound $\ell$-$\mathrm{CCS}(A^{3}\Pi)$ complex,
showing a conical intersection with $\ell$-$\mathrm{CCS}(X^{3}\Sigma^{-})$ at $R_{\mathrm{CS}}\!-\!R_{\mathrm{CC}}\!\approx\!+0.5\,{\AA}$~\citep{RIA003:15,TAR007:1129}. 
Differently from the ground $1^{3}A''$ state, the conversion from reactants to products 
as proceeding adiabatically through the $2^{3}A''$ and $1^{3}A'$ PESs entails a large activation barrier 
($\approx\!1\,\si{\electronvolt}$ at CASSCF/aV$T$Z level) which is located at
$R_{\mathrm{CS}}\!-\!R_{\mathrm{CC}}\!\approx\!-0.5\,{\AA}$; see Figure~\ref{fig:cas_c2}.
As shown, this region of the nuclear configuration space is extremely congested
by the existence of several low-lying excited triplet states 
correlating with $\mathrm{C_{2}}(a^{3}\Pi_{u})+\mathrm{S}(^{3}P)$. Note that 
the $\mathrm{C_{2}}(a^{3}\Pi_{u})+\mathrm{S}(^{3}P)$ reactants can approach
each other in six triplet ($3^{3}A''\!+\!3^{3}A'$), six singlet ($3^{1}A''\!+\!3^{1}A'$)
and six quintet ($3^{5}A''\!+\!3^{5}A'$) electronic states. For completeness,
their corresponding optimized reaction paths towards CS+C formation are also plotted
in Figure~\ref{fig:cas_c2}; see bottom, middle and top panels therein. 
Accordingly, when proceeding adiabatically, the reactions involving
$\mathrm{C_{2}}(a^{3}\Pi_{u})+\mathrm{S}(^{3}P)$ are all endothermic, leading ultimately to excited-state CS+C products. 
So, they are expected to be highly inefficient at the low temperature regimes here envisaged 
(unless non-adiabatic transitions play a role) and, for this reason, will 
not be considered further in this work.

Keeping now our focus on the target
$\mathrm{C_{2}}(X^{1}\Sigma_{g}^{+})+\mathrm{S}(^{3}P)\rightarrow\mathrm{CS}(X^{1}\Sigma^{+})+\mathrm{C}(^{3}P)$
process [Reaction ~(\ref{eq:c2_s})] and to better estimate its overall attributes,
we have performed MRCI+Q/aV$Q$Z//CASSCF/aV$Q$Z calculations along the underlying reaction paths
on both $1^{3}A''$, $2^{3}A''$ and $1^{3}A'$ electronic states.
The results are plotted in Figure~\ref{fig:mrci_c2}. Accordingly, at this level of theory, our best estimate for the  
exothermicity of reaction~(\ref{eq:c2_s}) is $1.05\,\si{\electronvolt}$ (without zero-point energy),
a value that  matches nearly perfectly the corresponding experimental
estimate of $1.04\,\si{\electronvolt}$~\citep{VIS019:1645,RED003:419}. The stabilization energies of the 
$\ell$-$\mathrm{CCS}(X^{3}\Sigma^{-})$ and $\ell$-$\mathrm{CCS}(A^{3}\Pi)$ complexes
are herein predicted to be $-5.3$ and $-4.2\,\si{\electronvolt}$, respectively, 
relative to the $\mathrm{C_{2}}(X^{1}\Sigma_{g}^{+})+\mathrm{S}(^{3}P)$ reactant channel. 
Most notably, Figure~\ref{fig:mrci_c2} (bottom panel) shows that, at MRCI+Q/aV$Q$Z//CASSCF/aV$Q$Z level, 
the predicted activation barriers along linear $C_{\infty v}$ paths for the $2^{3}A''$ and $1^{3}A'$ states are 
largely reduced with respect to CASSCF/aV$T$Z values (Figure~\ref{fig:cas_c2}; bottom panel), 
going from $\approx\!1$ to less than $0.1\,\si{\electronvolt}$. Indeed, 
by allowing the valence C--C--S angle ($\gamma_{\mathrm{CCS}}$) to also be freely optimized
in the MRCI+Q/aV$Q$Z//CASSCF/aV$Q$Z calculations, Figure~\ref{fig:mrci_c2} (top panel)
unequivocally pinpoints that such barriers actually become submerged,
thence lying below the corresponding $\mathrm{C_{2}}(X^{1}\Sigma_{g}^{+})+\mathrm{S}(^{3}P)$ reactant channel; note therein 
the existence of small discontinuities on the \emph{ab initio} 
curves which are associated with  
abrupt changes in $\gamma_{\mathrm{CCS}}$ near the top of these barriers.  
{This clearly indicates that all such PESs ($1^{3}A''$, $2^{3}A''$ and $1^{3}A'$)
  contribute to the overall dynamics/kinetics of reaction~(\ref{eq:c2_s}), even at low temperatures.
  In the following, we describe the methodology employed to obtain
  rate coefficients for this target reaction.}  

\subsection{Long range interactions}\label{subsec:longrange_c2}
%
%
Restricting the calculations to the equilibrium distance of C$_2$($^1\Sigma_g^+$), allow us to consider
C$_2$($^1\Sigma_g^+$) and  C$_2$($^3\Pi$) separately. Under this approximation, the situation simplifies
to a closed shell diatom plus an open shell atom  for the two asymptotes: C$_2$+S and CS+C. For both cases let us adopt the Jacobi coordinate system and expansion in terms of orthogonal functions 
 as defined eg. in \cite{Flower1977,Dubernet-Hutson:94}: 
\begin{equation}
	V(R,\theta,\phi,\theta_a,\phi_a)=
	\sum_{\lambda \lambda_a \mu} 
	V_{\lambda \lambda_a \mu}(R) C_{\lambda \mu}(\theta ,\phi) C_{\lambda_a, -\mu} (\theta_a \phi_a)
	\label{Vexpan}
\end{equation}
where $C_{\lambda \mu}(\theta, \phi) = \left(\frac{4 \pi}{2\lambda +1} \right)^{1/2} Y_{\lambda \mu}(\theta, \phi)$
are spherical harmonics in Racah normalization. The 
$\theta,\phi$  angles correspond to the orientation of diatomic molecules with respect to vector connecting the center-of mass of
the diatomic molecule with the atom in a Jacobi frame, while 
$\theta_a,\phi_a$  angles to the orientation of doubly occupied $p$ orbital of S and C atom, respectively. The $R$ is the  distance between the COM of diatom and atom. 
For both systems in $C_s$ symmetry the solution of electronic Schr{\"o}dinger equation is hard, since one cannot use single-reference methods since two solutions  will always belong to the same irreducible representation, and are nearly degenerate. The situation is simpler in case of symmetric configurations:  C$_{\infty v}$ (linear) - for both systems, and C$_{2v}$ (T-shaped) for   C$_2$+S.
For these cases atom+diatom states belong to distinct irreducible representations: for linear geometry  $\Pi$ and $\Sigma^-$ states, while for  C$_{2v}$ these symmetries are $B_1$, $B_2$ and $A_2$.
For these particular configurations of the CCS system
  one can use single-reference gold-standard CCSD(T) method for calculations of interaction energy in each  symmetry~\citep{Atahan2006,Alexander1998,Klos2004}.

Using linear and T-shape geometries we have calculated  UCCSD(T) interaction energies
in the range of 6-30 $\AA$ and converted it to 
$V_{\lambda \lambda_r \mu}(R)$ potentials using the following formula for C$_2$+S 
\begin{eqnarray}
	V_{000} &=& (2V_{\Pi} + V_{\Sigma} + 2V_{B_1} + 2V_{B_2} + 2 V_{A_2})/9 \\ 
	V_{020} &=& -2( V_{\Pi} - V_{\Sigma} + V_{B_1} + V_{B_2} - 2  V_{A_2})/9 \\ 
	V_{200} &=& 2(2V_{\Pi} + V_{\Sigma} -  2V_{B_1} - V_{B_2} -  V_{A_2})/9 \\ 
	V_{220} &=& -2(2V_{\Pi} - 2 V_{\Sigma} - V_{B_1} - V_{B_2} + 2 V_{A_2})/9 
\end{eqnarray}
and CS+C
\begin{eqnarray}  
	V_{000} &=& (2V_{\Pi,0} +  V_{\Sigma,0}  + 2V_{\Pi,180} + V_{\Sigma,180}  )/6  \\
	V_{100} &=& (2V_{\Pi,0} +  V_{\Sigma,0} - (2V_{\Pi,180} + V_{\Sigma,180} ))/6  \\ 
	V_{020} &=& ( V_{\Sigma,0} -  V_{\Pi,0} + V_{\Sigma,180} - V_{\Pi,180} )/3   \\ 
	V_{120} &=& ( V_{\Sigma,0} -  V_{\Pi,0} - (V_{\Sigma,180} - V_{\Pi,180} ))/3 
\end{eqnarray}
where in the latter equation $V_{\Pi,0}$/$V_{\Sigma,0}$  denote potential
for appropriate symmetry in  C-S-C  configuration and 
$V_{\Pi,180}$/$V_{\Sigma,180}$ in C-C-S alignment. The above equations can be obtained
by calculating Eq. \ref{Vexpan} for $\theta,\phi,\theta_a,\phi_a$
corresponding to symmetric configurations~\citep{Flower1977}. 

Then, the $V_{\lambda \lambda_r \mu}(R)$ potentials
were carefully fitted to analytical form to get inverse power expansion form.
It is important to realize, that collision-induced rotations of CS molecules are driven directly
by $V_{100}$ and $V_{120}$ terms, 
while for C$_2$ molecules colliding with atoms such terms are  $V_{200}$ and $V_{220}$.
When the atoms are assumed to be spherically symmetric,
the terms $V_{120}$ and $V_{220}$ can be ignored.
Thus, from now on we will skip the dependence on $\theta_a,\phi_a$ in the  Eq. \ref{Vexpan}.
Thus our model of the potential used  for the statistical method includes only isotropic term $V_{000}$
and leading anisotropies
$V_{100}$ and $V_{200}$ which were fitted to analytical forms of van der Waals expansion:
$\sum_{i=0}^3 C_{6+i} R^{-(6+i)}$ for $V_{000}$ and $V_{200}$, and   $\sum_{i=0}^3 C_{7+i} R^{-(7+i)}$ for $V_{100}$.
These potentials can be viewed as averaged over all orientations of $P$-state atoms. 
Moreover, for the leading coefficients we also performed similar calculations using open-shell Symmetry Adapted Perturbation theory \citep{Hapka2012} and confirmed the values for the leading coefficients $C_6$ and $C_7$
of the reactants and products  (they agreed to within 10\%).

The final analytical form of the potential used for C$_2$+S reads (in atomic units of distance and energy)
\begin{eqnarray}
  V(R,\theta) &=&  A \, R^{-6}  +B\,   R^{-8}  + C\, R^{-10} \\
   &+&   \frac{1}{2}(3 \cos^2(\theta) -1)
  ( D\, R^{-6} + E\, R^{-8} + F\,  R^{-10} ) \nonumber
\end{eqnarray}
with A=$-125.8$ , B=$-9.444\times 10^3  $,C=$1.743\times 10^5$, D=$-13.66$ , E=$-2.848\times 10^3 $ and F=$- 1.929\times 10^5 $  
while for CS+C
\begin{eqnarray}
  V(R,\theta) &=&  A\,  R^{-6}  +B\,  R^{-8}  + C\, R^{-10} \\
              &+& D\, R^{-12} + {\cos( \theta)\over R^{-7}} \,( E\, + F\,  R^{-2}  + G\,  R^{-4} +H\, R^{-6}) \nonumber
\end{eqnarray}
with A= $- 147.5$, B=$- 207.3$, C=$- 3.834\times 10^{6}$, D=$-2.805 \times 10^{8}$,
E=$- 487.3$, F=$-4.309\times 10^{3} $, G=$-2.394\times 10^{7}$
and H=$ - 1.997\times 10^{9} $.

\begin{figure}
\centering
\includegraphics[angle=0,width=1\linewidth]{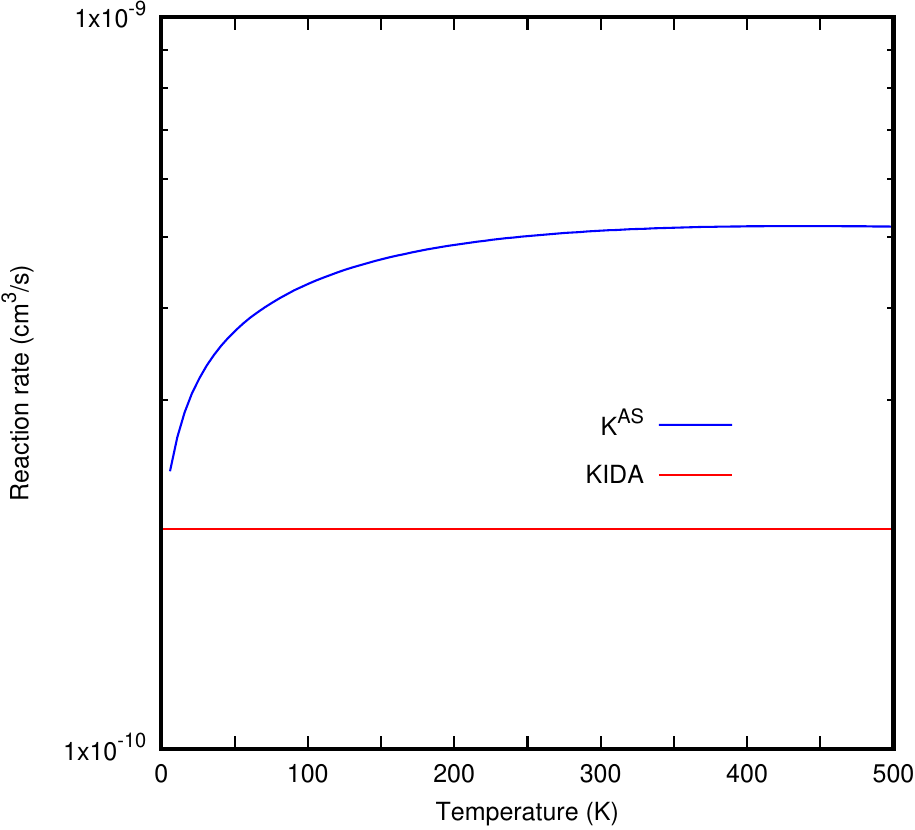}
\caption{\footnotesize Reactive rate constant for the C$_2$+S $\rightarrow$ CS + S collisions obtained with the present AS method
and compared to that available in the KIDA data base from  \cite{Vidal-etal:17}} 
\label{fig:rateC2+S}
\end{figure}

\begin{table}[hbtp]
\centering
\caption{\label{tab:new-rates}Parameters used to fit the total reaction rates
calculated for CH+S and C$_2$ + S, according to the expression $K(T) = A (T/300)^B e^{-C/T}$  }
\begin{tabular}{|c | c c c|}
  \hline
  reaction  & A (cm$^3/s$) & B & C (K)  \\
  \hline
  CH + S & 7.52\ $\times$ 10$^{-11}$ & -0.117 & 5.42 \\
  C$_2$ + S & 5.08\ $\times$ 10$^{-10}$ & 0.121 & 2.70\\
  \hline
\end{tabular}
\end{table}
\subsection{Rate constant calculations}\label{subsec:rate_c2}

Due to  the deep well appearing for the three lowest adiabatic states
of C$_2$+ S reaction, and the large masses of the three atoms involved, this system
has a high density of resonances near the thresholds. Therefore,
this reaction is expected to be governed by a statistical mechanism.
In this work we shall use the Adiabatic Statistical (AS) method \citep{Quack-Troe:74},
using the AZTICC code recently implemented \citep{Gomez-Carrasco-etal:22}.
Here, we shall use the rigid rotor approach, similar
to the method already used to treat several reactions and inelastic processes \citep{Konings-etal:21}.
We consider the experimental exoergic values of D$_0$=D$^{C_2}_0$ - D$^{CS}_0$ = 1.14 eV
\citep{Herzberg-etal:79}, which already
includes the vibrational zero-point energy (ZPE) of reactants and products. This exothermicity
is very close to that shown in the {\it ab initio} calculations, of $\approx$ 1. eV  in Fig.~\ref{fig:mrci_c2}, which does
not include ZPE.

In the rigid rotor approach, the diatomic distances for the diatomic molecules are frozen
to their equilibrium values, $r_e$= 1.2425 and 1.568 \AA\ for C$_2$ and CS, respectively.
Since C$_2$ is homonuclear, the even and odd rotational channels are not coupled, and are treated independently.
The large exothermicity requires the inclusion of vibrational levels for the CS products, up to v=15, to describe
adequately the products density of states. To this end, the adiabatic potentials are generated
for each final $v$, independently,
using the rigid rotor approximation at $r_{eq}$ with the same long-range potential
but shifting the energy using the anharmonic
vibration constants, $\omega_e=$ 0.15933 eV $\omega_e\chi_e=$ 0.801 meV  from \cite{Herzberg-diatomics}.

\begin{table*}
\caption{Model predictions using the old and new reaction rate coefficients}
\label{Table:CS}
\begin{tabular}{llllllll}
\hline 
Molecule & Age (Myr)    &   TMC1-old     &  TMC1-new  & HC-old  &  HC-new   &  PDR-old   & PDR-new  \\ \hline
CS  &   0.1  & 1.92$\times 10^{-8}$  & 1.61$\times 10^{-8}$  & 6.33$\times 10^{-11}$   & 6.13$\times 10^{-11}$   & 2.69$\times 10^{-9}$   & 2.88$\times 10^{-9}$   \\
       &  1.0  & 6.47$\times 10^{-8}$  & 6.47$\times 10^{-8}$  & 2.04$\times 10^{-12}$  & 1.66$\times 10^{-12}$  & 2.89$\times 10^{-9}$  & 3.06$\times 10^{-9}$  \\
SO  &  0.1  & 1.08$\times 10^{-7}$  & 4.70$\times 10^{-8}$  & 6.36$\times 10^{-7}$   & 6.28$\times 10^{-7}$   & 1.70$\times 10^{-9}$   & 1.70$\times 10^{-9}$   \\
       & 1.0  & 1.70$\times 10^{-9}$  & 1.67$\times 10^{-9}$  & 4.28$\times 10^{-7}$   & 3.50$\times 10^{-7}$   & 1.82$\times 10^{-9}$  & 1.82$\times 10^{-9}$  \\
SO$_2$  & 0.1  & 7.68$\times 10^{-9}$   & 1.41$\times 10^{-8}$     & 1.32$\times 10^{-7}$   & 1.56$\times 10^{-7}$   & 9.99$\times 10^{-14}$   & 1.33$\times 10^{-13}$   \\
               & 1.0  & 4.28$\times 10^{-11}$  & 2.20$\times 10^{-10}$  & 3.71$\times 10^{-7}$   & 4.49$\times 10^{-7}$   & 1.06$\times 10^{-13}$   & 1.43$\times 10^{-13}$  \\
\hline
\end{tabular}
\end{table*}

All product ro-vibrational levels
are then considered to obtain the square of the S-matrix at each total angular momentum $J$
(see Ref.\cite{Gomez-Carrasco-etal:22} for more details).
The calculations are done using a body-fixed frame, in which the z-axis is parallel to the Jacobi vector {\bf R}, joining
the diatomic center-of-mass to the atom, and the three atoms are considered in the body-fixed xz-plane.
In the present calculations, a maximum rotational quantum number $j_{max}=$ 200 and 250 have been considered
for C$_2$ and CS, and a maximum helicity quantum number of $\Omega_{max}$= 15.
The individual state-to-state reactive cross sections are obtained performing the summation over all $J$ in the partial wave expression up to $J_{max}=$ 200.

Finally, integrating over the translational
energy according to a Boltzmann energy distribution, and summing over all accessible states, the thermal rate
constants, $K_\alpha(T)$,  are obtained, with $\alpha$= $1^3A''$, $1^3A'$and  $1^3A'$.
The $2^3A''$ and $1^3A'$ state present a submerged barrier
which could reduce the reactivity, but 
here we consider that the
three $\alpha$ electronic states have the same reactive rate constant.
For this reason, when the spin-orbit splitting of S($^3P_J$) is included, the final thermal rate constant
averaging over the spin-orbit states, $K^{AS}(T)$, is equal to  $K_\alpha(T)$,
shown in Fig.~\ref{fig:rateC2+S},
whose parameters are fitted and listed in Table~\ref{tab:new-rates}.

The C$_2$+ S $\rightarrow$ CS + S available in KIDA data base\footnote{https://kida.astrochem-tools.org/}
is that of \cite{Vidal-etal:17} obtained by a capture
method, which corresponds to a value
of $K^{C}(T)$= 2\, 10$^{-10}$ cm$^3$/s, independent of temperature, and was calculated using a capture method,
$i.e.$ assuming that the reaction
is exothermic, but without considering the deep well described in this work. Such approach neglects the possibility
that the  C$_2$S complex formed under the statistical assumption can exit back to C$_2$ + S products. This probability
is small because it is proportional to the density of states in each channel, and this density is much lower in the C$_2$+S
due to the exothermicity, the homonuclear symmetry and the larger rotational constant. 

\section{Discussion}

In this section we evaluate the impact of the new reaction rates in our understanding of interstellar chemistry.
This is not straightforward since the formation 
and destruction routes of the different species depend on the local physical and
chemical conditions as well as the chemical time. Therefore, we need to consider different environments
to be able to have a comprehensive view
of the impact of the new reactions rates on astrochemical calculations,
and in particular in our ability to reproduce the abundance of CS.   

We performed chemical calculations using Nautilus 1.1 \citep{Ruaud2016},
a three-phase model, in which gas, grain surface and grain mantle phases, and their interactions, are considered.
We used the code upgraded as described by \citet{Wakelam2021} with the chemical network
of KIDA$^2$  that
has been modified to
account for the reaction rates estimated by \citet{Fuente2016}, \citet{Fuente2019},
and this paper {(reaction rates in Table ~\ref{tab:new-rates})}. 
In Nautilus, desorption into the gas phase is only allowed for the surface species, considering both thermal and non-thermal mechanisms.
In the regions where the temperature of grain particles is below the sublimation temperature,
non-thermal desorption processes become important to calculate the number of
molecules in gas phase. The latter include desorption induced by cosmic rays \citep{Hasegawa1993},
direct (UV field) and indirect (secondary UV field induced by the cosmic-ray flux) photo-desorption,
and reactive chemical desorption \citep{Garrod2007, Minissale2016}. In the following calculations,
we use the prescription proposed by \citet{Minissale2016} for ice coated grains to calculate the reactive chemical desorption.
The physical and chemical conditions associated with these three simulations are detailed below:

\begin{table*}
\caption{Chemical abundances in TMC 1 (CP)}
\label{Table:obs}
\begin{tabular}{lllll}
\hline 
\multicolumn{1}{l}{Species}                    &
\multicolumn{2}{c}{Observed}                &
\multicolumn{1}{l}{Model (0.1 Myr)} & 
\multicolumn{1}{l}{Model (1 Myr)} \\ \hline
H$_2$    &                   0.5                     &                                          &                                                                    &                \\
CS             &  6.5$\times 10^{-9}$$^{(1)}$   & 6.4$\times 10^{-9}$$^{(2)}$              & 1.6$\times 10^{-8}$   &  6.5$\times 10^{-8}$  \\
CO            &  4.8$\times 10^{-5}$$^{(1)}$   & 4.2$\times 10^{-5}$$^{(2)}$               & 1.0$\times 10^{-4}$   &  6.8$\times 10^{-6}$ \\
HCO$^+$  &  5.0$\times 10^{-9}$$^{(1)}$   & 8.4$\times 10^{-9}$$^{(2)}$               & 1.0$\times 10^{-8}$   &  2.0$\times 10^{-9}$  \\
HCS$^+$  &  1.1$\times 10^{-10}$$^{(*)}$  & 7.0$\times 10^{-11}$$^{(1)}$             &  2.3$\times 10^{-11}$ & 1.4$\times 10^{-10}$ \\
HCN         &  4.8$\times 10^{-8}$$^{(*)}$     &  3.5$\times 10^{-8}$$^{(2)}$             &  5.0$\times 10^{-8}$    &  2.1$\times 10^{-7}$ \\ 
SO            &  9.0$\times 10^{-10}$$^{(1)}$  & 5.3$\times 10^{-10}$$^{(2)}$            & 4.7$\times 10^{-8}$    &  1.7$\times 10^{-9}$ \\
OCS         &  1.2$\times 10^{-10}$$^{(2)}$   &                                                          &  3.2$\times 10^{-9}$    &  4.5$\times 10^{-9}$ \\
H$_2$S   &  8.0$\times 10^{-10}$$^{(2)}$    & 1.1$\times 10^{-9}$$^{(3)}$              & 1.4$\times 10^{-9}$    &  1.3$\times 10^{-8}$ \\
H$_2$CS  &  1.1$\times 10^{-9}$$^{(4)}$    &                                                          &  2.1$\times 10^{-9}$    &  6.7$\times 10^{-8}$ \\
C$_2$S       &  2.8$\times 10^{-9}$$^{(4)}$  &                                                          &  2.6$\times 10^{-10}$   &  9.4$\times 10^{-9}$ \\
C$_3$S      &  3.8$\times 10^{-10}$$^{(4)}$ &                                                          &  3.9$\times 10^{-10}$   &  3.3$\times 10^{-9}$ \\
HCSCN       &  8.9$\times 10^{-12}$$^{(5)}$ &                                                         &                                       &                    \\
HCSCCH     &  2.2$\times 10^{-12}$$^{(5)}$ &                                                         &                                       &                    \\
NCS             &  2.2$\times 10^{-11}$$^{(5)}$ &                                                         &                                       &                      \\
HCCS          &  1.9$\times 10^{-11}$$^{(5)}$ &                                                         &  1.0$\times 10^{-11}$     & 1.1$\times 10^{-8}$ \\
H$_2$CCS    & 2.2$\times 10^{-11}$$^{(5)}$ &                                                        &  2.1$\times 10^{-11}$     & 4.7$\times 10^{-11}$  \\
H$_2$CCCS  & 1.0$\times 10^{-11}$$^{(5)}$ &                                                       &                                        &                     \\
C$_4$S         & 1.0$\times 10^{-12}$$^{(5)}$ &                                                        &  6.2$\times 10^{-12}$     & 3.1$\times 10^{-10}$ \\
C$_5$S         & 1.4$\times 10^{-12}$$^{(5)}$ &                                                        &                                         &                    \\
HC$_3$S$^+$   & 5.5$\times 10^{-12}$$^{(5)}$ &                                                   &  9.3$\times 10^{-12}$       & 6.5$\times 10^{-11}$ \\ 
HSCN                & 1.9$\times 10^{-12}$$^{(6)}$ &                                                   &  6.3$\times 10^{-13}$       & 1.3$\times 10^{-10}$  \\ 
HNCS               &  1.9$\times 10^{-12}$$^{(6)}$ &                                                   &  8.8$\times 10^{-13}$       & 2.0$\times 10^{-10}$  \\ 
HCCCCS         &  2.6$\times 10^{-12}$$^{(7)}$ &                                                    &                                          &                    \\
HCCS$^+$      &  3.0$\times 10^{-11}$$^{(8)}$ &                                                    &  1.4$\times 10^{-11}$        & 1.3$\times 10^{-10}$  \\
NS$^+$           &  1.4$\times 10^{-12}$$^{(9)}$ &                                                    &  1.7$\times 10^{-11}$        & 1.5$\times 10^{-12}$   \\  
NS                  &  4.7$\times 10^{-11}$$^{(9)}$ &                                                     &  2.2$\times 10^{-10}$        & 1.0$\times 10^{-9}$      \\
HCS               &  1.5$\times 10^{-10}$$^{(5)}$ &                                                     &  2.1$\times 10^{-11}$        & 3.1$\times 10^{-10}$   \\ 
\hline
\end{tabular}

\noindent
References: (1) \citet{Fuente2019} 2) \citet{Rodriguez-Baras2021} 3) \citet{Navarro2020}
4) \citet{Gratier2016} 5) \citet{Cernicharo2021b} 6) \citet{Adande2010} 7) \citet{Fuentetaja2022}
8) \citet{Cernicharo2021a} 9) \citet{Cernicharo2018}

\noindent
(*) Calculated in this work.

\end{table*}

\begin{itemize}
\item TMC~1: This case represents the physical and chemical conditions prevailing in molecular
  cloud complexes where low-mass stars are formed.
  The physical properties of these regions are typically described with moderate
  number densities of atomic hydrogen nuclei n$_{\rm H}=3\times 10^4$ cm$^{-3}$,
  cold gas and dust temperatures ${\rm T}= 10$ K,
  a moderate visual extinction ${\rm A}_{V}=20$ mag,
  a cosmic-ray H$_{2}$ ionization rate of $\zeta_{{\rm H}_{2}}=10^{-16}$ s$^{-1}$ {\citep{Fuente2019,  Fuente2023}},
  and an intensity of the far-ultraviolet field (FUV) equal to $\chi=5$ in Draine units.
  We assume that  sulphur is depleted by a factor of 20 relative to cosmic abundance
  as derived by \citet{Fuente2019} and \citet{Fuente2023} in Taurus and Perseus.

\item Hot core: This case represents the physical and chemical conditions in the warm interior of young protostars where the gas and dust temperature is $>$100 K and the icy grains mantles are sublimated. Typical physical conditions in these regions are: n$_{\rm H}$ = 3 $\times$ 10$^6$ cm$^{-3}$, T$_k$= 200 K, A$_V$= 20 mag, $\zeta_{\rm H_2}$= 1.3 $\times$ 10$^{-17}$ s$^{-1}$,
  $\chi$=1 in Draine units. Sulphur depletion in hot cores is not well established. We adopted  [S/H]=8$\times$10$^{-7}$ since this value is commonly used to model massive hot cores \citep{Gerner2014}.

\item Photon-dominated regions (PDRs)  are those environments where the far-ultraviolet photons emitted by hot
stars are determining the physical and chemical conditions of gas and dust. PDRs can be found on the surfaces of protoplanetary disks
and molecular clouds, globules, planetary nebulae, and starburst galaxies. As representative of the physical and chemical conditions
in the PDRs associated with massive star forming regions, we select:  n$_{\rm H}$ =  5 $\times$ 10$^5$ cm$^{-3}$, T$_k$= 100 K, A$_V$= 4 mag,
$\zeta_{H_2}$= 10$^{-16}$ s$^{-1}$, $\chi$=10$^4$ in Draine units. The amount of sulphur in gas phase in PDRs is still an open question.
Based on observations of sulphur recombination lines in the Orion Bar, \citet{Goicoechea2021} obtained that the abundance of sulphur
should be close to the solar value in this prototypical PDR.
We are aware that sulphur recombination lines are arising from PDR layers close
to the S$^+$/S transition, at $A_v\sim$ 4 mag \citep{Simon1997,Goicoechea:2006})
and sulphur depletion might be higher towards more shielded regions from where the emission of most sulphur-bearing molecules comes.
In spite of this, we adopt [S/H]=1.5$\times$10$^{-5}$ for our calculations. 

\end{itemize}

\begin{figure*}
\centering
\includegraphics[angle=0,width=0.85\linewidth]{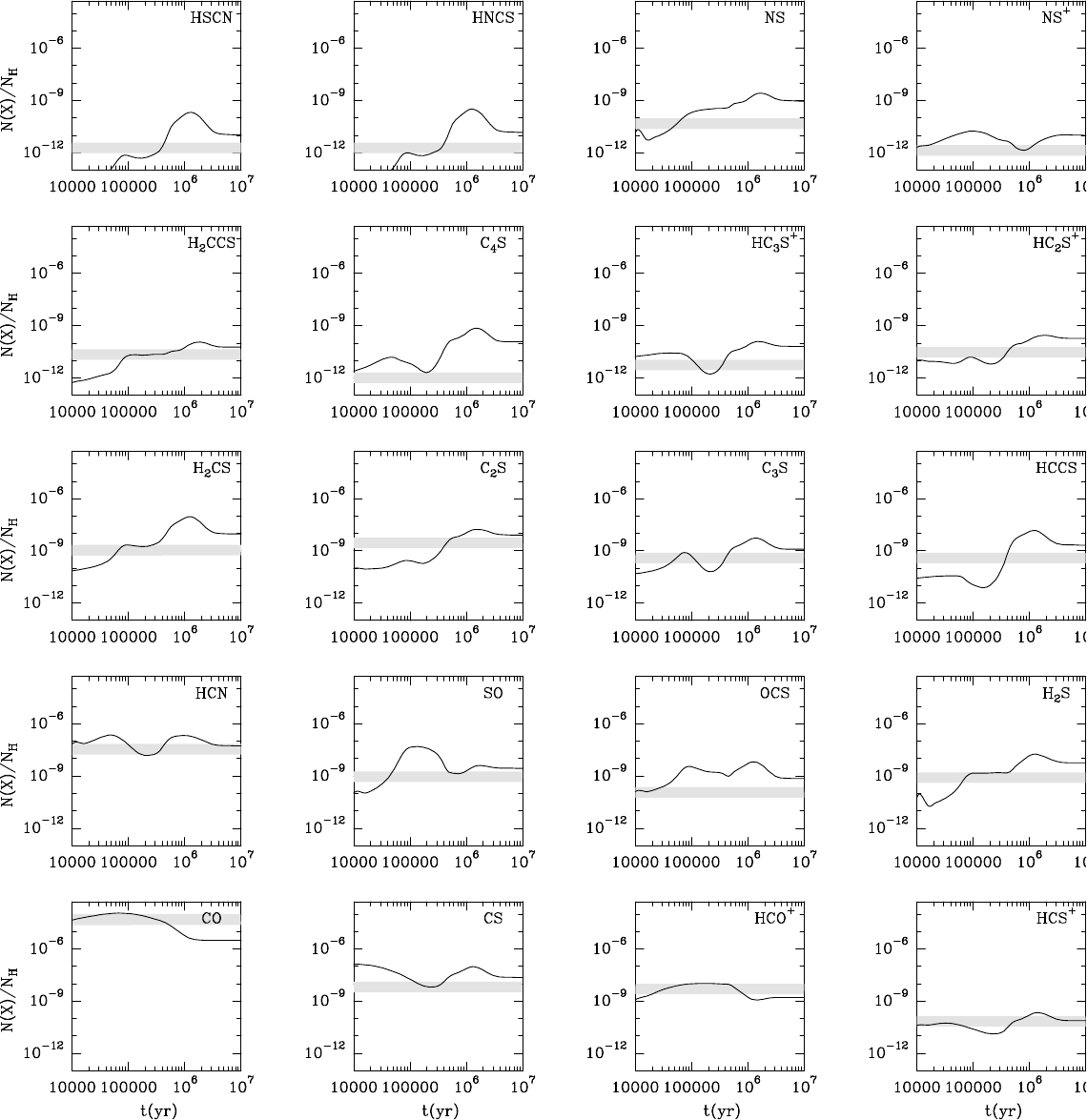}
\caption{Comparison between Nautilus predictions obtained using our updated chemical network and the abundances observed towards TMC 1 (CP)
  (see Table~\ref{Table:obs}). Grey regions indicate the observed values. We assumed an uncertainty of a factor of 2 in the measured abundances. The physical and chemical conditions used in our calculations are: n$_H$ = 3 $\times$ 10$^4$ cm$^{-3}$, T$_k$= 10 K, A$_V$= 20 mag, $\zeta_{H_2}$= 10$^{-16}$ s$^{-1}$, $G_{0}$=1 in Draine units and [S/H]=7.5$\times$10$^{-7}$ (sulphur depleted by a factor of $\sim$20).} 
\label{fig:chem}
\end{figure*}

Table~\ref{Table:CS} shows the fractional abundances of CS, SO, and SO$_2$ predicted using the "old" and "new" chemical network
and the physical conditions above described.
{In regions where the ionization fraction is large, CS is essentially produced from the electronic dissociative recombination
of HCS$^+$, where HCS$^+$ is formed by reactions of S$^+$ and CH \citep{SternbergDalgarno1995, LucasLiszt2002}.
 \citet{Vidal-etal:17} showed that the  production of HCS$^+$ with S$^+$  + CH $\rightarrow$ CS$^+$ + H, and CS$^+$ + H$_2$ $\rightarrow$ HCS$^+$ + H, is the most
efficient HCS$^+$ formation route at the cloud surface.}
In more shielded regions where the sulphur is mainly in neutral atomic form, CS is also produced by neutral-neutral reactions,
with significant contributions of the reactions studied in this paper. {For dark cloud conditions,  the reaction C$_2$ + S forms CS more efficiently than  CH + S.  At low temperatures, the calculated C$_2$ + S rate is slightly higher than previous value  (see Fig.~\ref{fig:rateC2+S}). However, its possible effect is canceled by the lower value 
of the new CH + S reaction rate (see Fig.~\ref{fig:CH+S-thermal-rate}). As a result, the impact of our new rates on the CS abundance for the TMC1 case is negligible. 
Not surprisingly, the major impact of the new reactions rates calculated on the CS abundance is observed for the hot core case with variations of $\sim$ 20\% due
to the significantly higher  C$_2$ + S reaction rate at temperatures $>$100~K (see Fig.~\ref{fig:rateC2+S}).
The S + O$_2$ $\rightarrow$ SO + O and  SO + OH $\rightarrow$ SO$_2$ + H reactions rates published by \citet{Fuente2016}
and \citet{Fuente2019} produce the maximum variations in the case TMC1, with variations
of the SO and SO$_2$ abundances of a factor of  $>$2, which demonstrates the need of performing this type of calculations.}


TMC 1 (CP) is the astrophysical object for which  the highest number of sulphur-bearing species have been detected so far,
with more than ten complex sulphur-bearing molecules detected for the first time in the last 3 years
(\citealp{Cernicharo2021b, Fuentetaja2022}, Table~\ref{Table:obs}).
The large number of atoms in these new species ($>$5) shows that
a rich and complex organo-sulphur chemistry is going on in this dark cloud\citep{Laas-Caselli:19}.
Although these large molecules carry a small percentage of the sulphur budget, their detection
is useful to test the predictive power of our chemical network. Table~\ref{Table:obs}
shows a compilation of the observed abundances towards TMC 1 (CP).
In order to perform the most uniform and reliable comparison,
the abundances have been re-calculated assuming N$_{\rm H}$=3.6$\times$10$^{22}$ cm$^{-2}$ \citep{Fuente2019}. The abundances of HCS$^+$ and
H$^{13}$CN were already estimated by \citet{Rodriguez-Baras2021}. Here, they have been re-calculated using the most recent collisional coefficients reported by
\citet{Denis2022} and \citet{Navarro2023}. Following the same methodology explained in \citet{Fuente2019} and 
\citet{Rodriguez-Baras2021}, we assumed X(HCN)/X(H$^{13}$CN)=60 to estimate the HCN abundance.

We performed chemical calculations using Nautilus 1.1 and the chemical network modified to
account for the reaction rates presented in this paper, \citet{Fuente2019}, and \citet{Fuente2016}, and
the physical conditions derived by \citet{Fuente2023}.These physical conditions are the same as in the TMC1 
case in Table~\ref{Table:CS}. { Only 17 of the total number of sulphur-bearing species
detected towards TMC 1 (CP) are included in our chemical network.}
Fig.~\ref{fig:chem} shows the chemical predictions for all the sulphur-bearing species that have been observed in this 
proto-typical source and are included in our chemical network.
In addition, we show the CO, HCO$^+$, and HCN abundances because these molecules are considered good
tracers of the gas ionization degree and C/O ratio \citep{Fuente2019}.

We find a reasonable agreement (within a factor of 10) between model and observations for
CO, HCO$^+$, HCN, CS, HCS$^+$, H$_2$S, SO, 
C$_2$S, C$_3$S, C$_4$S, H$_2$CS, HC$_2$S$^+$, HC$_3$S$^+$, H$_2$CCS, NS$^+$, HNCS, and HSCN
for times between 0.1~Myr and 1~Myr. However, as already commented by \citet{Bulut-etal:21} and \citet{Wakelam2021},
the chemical time at which we find the best solutions depends on the considered species.
The most important restrictions respect to the chemical time comes for CO and HCO$^+$
whose abundances rapidly decrease for times later than  0.4 Myr. On the opposite side,
we find that the abundances of HCS$^+$, SO, and C$_2$S are better reproduced
for times $>$ 1 Myr.  Only OCS, {C$_4$S}, and NS cannot be fitted with chemical times between 0.1 Myr and 1 Myr.
We would like to recall that the chemical time
is not the same as the dynamical time, since various physical phenomena such as turbulent motions
that carry molecules to the cloud surfaces or shocks can reset the chemical age of the gas.
sulphur-bearing species are very sensitive to the chemical time with several species whose abundances
vary in several orders of magnitude from 0.1 Myr to 1 Myr.
Therefore the chemical time is a critical parameter to fit them (see Fig.\ref{fig:chem}).
In our 0D chemical calculations, the physical conditions remain fixed, which is far
from the real case of a collapsing and fragmenting cloud. In forthcoming papers,
we will explore the influence that the cloud dynamical evolution would have on the sulphur chemistry. 

\section{Summary and conclusions}

The rate constants for the formation of  CS($X^1\Sigma^+$),  CH$(^2\Pi)$ + S($^3P$)
and C$_2$($X^1\Sigma_g^+$)+ S($^3P$) have been obtained in this work, and are tabulated in Table~\ref{tab:new-rates}. These two reactions
involve open shell reactants, and therefore present several degenerate, or nearly degenerate, electronic states. The role
of each initial electronic state in the formation of CS  has been analyzed in detail.

For  CH$(^2\Pi)$ + S($^3P$)
it is found that only the 1$^2A'$ can contribute to the CS($X^1\Sigma^+$) formation through an exothermic
barrierless mechanism, $i.e.$ 2 states of the doublet
among the 36 degenerate electronic states correlating to the  CH$(^2\Pi)$ + S($^3P$) asymptote. When the spin-orbit
splitting is taken into account, at the low temperatures of 10 K the electronic partition function
becomes 2/9, tending to 2/36 at high temperature. Surprisingly,
the rate constant obtained with a capture model \citep{Vidal-etal:17}
is only a factor of two higher than the present result at 10 K, and this difference increases
very little with increasing temperature up to 500K.

For the C$_2$($X^1\Sigma_g^+$)+ S($^3P$) reaction the three triply degenerate states connect to the  CS($X^1\Sigma^+$) products.
It is found that the three states present a deep insertion well, with depths between 5.5 and 4 eV. The ground
electronic state proceeds with no barrier, while the two excited states have a barrier in the products channel, which
becomes submerged for bent configuration. The presence of the deep insertion well justifies the use of an adiabatic
statistical method to calculate the reactive rate constant, which in turn is very similar to
that obtained with a capture model \citep{Vidal-etal:17} at 10 K. The difference increases with temperature, and
the present results become 2.5 times larger than the constant  value obtained 2 $\times$10$^{-10}$ cm$^3$/s at 500 K.

The present results corroborate those obtained with classical capture models
at low temperatures, differing by a factor of 2.5 at most. It should be noted, however,
that this is due to different reasons for the two reactions studied here. For
open shell reactants, as those treated here, for which experiments are difficult, it is
required to do a detailed analysis of the reactivity of all initial degenerate electronic states
of the reactants before generalizing these findings. 

The new rates have been implemented in a chemical network to compare 
with the observations of sulphur-bearing species towards
TMC 1(CP). Model predictions are in reasonable agreement with the observation for most the sulphur-bearing species,
except for OCS and NS, which cannot be fitted with our model. 
However, it is not possible to fit all of them with an unique chemical time, which suggests that dynamical
effects are important for sulphur chemistry.

\begin{acknowledgements}
 The research leading to these results has received funding from
 MICIN (Spain) under grant PID2021-122549NB-C21 and from the 
 {European Union's Horizon 2020 research and innovation program under the Marie Sklodowska-Curie} grant agreement No 894321.
 N.B is grateful for support from the Polish National Agency for Academic Exchange (NAWA) Grant and
 also acknowledges TUBITAK's 2219-Program by scholarship no. 1059B192200348.
 P.S.Z. is grateful to National Science Centre of Poland for funding the project No 2019/34/E/ST4/00407.
 AF and PRM are grateful to Spanish MICIN for funding under grant PID2019-106235GB-I00.
 JRG thanks the Spanish MCINN for funding support under grant PID2019-106110GB-I00.
 JEP was supported by the Max-Planck Society. DNA acknowledges funding support from
 Fundaci\'on Ram\'on Areces through its international postdoc grant program.
 RLG would like to thank the "Physique Chimie du Milieu Interstellaire" (PCMI)
 programs of CNRS/INSU for their financial supports.
 
\end{acknowledgements}


\begin{thebibliography}{90}
\expandafter\ifx\csname natexlab\endcsname\relax\def\natexlab#1{#1}\fi

\bibitem[{{Adande} {et~al.}(2010){Adande}, {Halfen}, {Ziurys}, {Quan}, \&
  {Herbst}}]{Adande2010}
{Adande}, G.~R., {Halfen}, D.~T., {Ziurys}, L.~M., {Quan}, D., \& {Herbst}, E.
  2010, \apj, 725, 561

\bibitem[{Aguado \& Paniagua(1992)}]{Aguado-Paniagua:92}
Aguado, A. \& Paniagua, M. 1992, J. Chem. Phys., 96, 1265

\bibitem[{Aguado {et~al.}(1998)Aguado, Tablero, \& Paniagua}]{Aguado-etal:98}
Aguado, A., Tablero, C., \& Paniagua, M. 1998, Comput. Phys. Commun., 108, 259

\bibitem[{{Ag{\'u}ndez} \& {Wakelam}(2013)}]{Agundez2013}
{Ag{\'u}ndez}, M. \& {Wakelam}, V. 2013, Chemical Reviews, 113, 8710

\bibitem[{Alexander(1998)}]{Alexander1998}
Alexander, M.~H. 1998, The Journal of chemical physics, 108, 4467

\bibitem[{Atahan {et~al.}(2006)Atahan, K{\l}os, Zuchowski, \&
  Alexander}]{Atahan2006}
Atahan, S., K{\l}os, J., Zuchowski, P.~S., \& Alexander, M.~H. 2006, Physical
  Chemistry Chemical Physics, 8, 4420

\bibitem[{{Boogert} {et~al.}(1997){Boogert}, {Schutte}, {Helmich}, {Tielens},
  \& {Wooden}}]{Boogert1997}
{Boogert}, A.~C.~A., {Schutte}, W.~A., {Helmich}, F.~P., {Tielens},
  A.~G.~G.~M., \& {Wooden}, D.~H. 1997, \aap, 317, 929

\bibitem[{Bulut {et~al.}(2021)Bulut, Roncero, Aguado, Loison, Navarro-Almaida,
  Wakelam, Fuente, end R~Le~Gal, Caselli, Gerin, Hickson, Spezzano,
  Rivi\'ere-Marichalar, Alonso-Albi, Bachiller, Jimenez-Serra, Kramer, Tercero,
  Rodriguez-Baras, Garc{\'\i}a-Burillo, Goicoechea, no~Morales, Esplugues,
  Cazaux, Commercon, Laas, Kirk, Lattanzi, Mart{\'\i}n-Dom\'nech, noz Caro,
  Pineda, Ward-Thompson, Marcelino, Malinen, Friesen, Giuliano, Agxi\'undez, \&
  Hacar}]{Bulut-etal:21}
Bulut, N., Roncero, O., Aguado, A., {et~al.} 2021, Astron. AstroPhys., 646, A5

\bibitem[{{Cazaux} {et~al.}(2022){Cazaux}, {Carrascosa}, {Mu{\~n}oz Caro},
  {Caselli}, {Fuente}, {Navarro-Almaida}, \&
  {Rivi{\'e}re-Marichalar}}]{Cazaux2022}
{Cazaux}, S., {Carrascosa}, H., {Mu{\~n}oz Caro}, G.~M., {et~al.} 2022, \aap,
  657, A100

\bibitem[{{Cernicharo} {et~al.}(2021{\natexlab{a}}){Cernicharo}, {Cabezas},
  {Endo}, {Ag{\'u}ndez}, {Tercero}, {Pardo}, {Marcelino}, \& {de
  Vicente}}]{Cernicharo2021b}
{Cernicharo}, J., {Cabezas}, C., {Endo}, Y., {et~al.} 2021{\natexlab{a}}, \aap,
  650, L14

\bibitem[{{Cernicharo} {et~al.}(2021{\natexlab{b}}){Cernicharo}, {Cabezas},
  {Endo}, {Marcelino}, {Ag{\'u}ndez}, {Tercero}, {Gallego}, \& {de
  Vicente}}]{Cernicharo2021a}
{Cernicharo}, J., {Cabezas}, C., {Endo}, Y., {et~al.} 2021{\natexlab{b}}, \aap,
  646, L3

\bibitem[{{Cernicharo} {et~al.}(2018){Cernicharo}, {Lefloch}, {Ag{\'u}ndez},
  {Bailleux}, {Margul{\`e}s}, {Roueff}, {Bachiller}, {Marcelino}, {Tercero},
  {Vastel}, \& {Caux}}]{Cernicharo2018}
{Cernicharo}, J., {Lefloch}, B., {Ag{\'u}ndez}, M., {et~al.} 2018, \apjl, 853,
  L22

\bibitem[{Davidson(1975)}]{Davidson:75}
Davidson, E.~R. 1975, J. Comp. Phys., 17, 87

\bibitem[{{Denis-Alpizar} {et~al.}(2022){Denis-Alpizar}, {Quintas-S{\'a}nchez},
  \& {Dawes}}]{Denis2022}
{Denis-Alpizar}, O., {Quintas-S{\'a}nchez}, E., \& {Dawes}, R. 2022, \mnras,
  512, 5546

\bibitem[{Dubernet \& Hutson(1994)}]{Dubernet-Hutson:94}
Dubernet, M.~L. \& Hutson, J. 1994, J. Chem. Phys., 101, 1939

\bibitem[{Dunning(1989)}]{DUN89:1007}
Dunning, T.~H. 1989, J. Chem. Phys., 90, 1007

\bibitem[{Dunning \& Jr.(1989)}]{Dunning:89}
Dunning, T.~H. \& Jr. 1989, J. Chem. Phys., 90, 1007

\bibitem[{{Esplugues} {et~al.}(2022){Esplugues}, {Fuente}, {Navarro-Almaida},
  {Rodr{\'\i}guez-Baras}, {Majumdar}, {Caselli}, {Wakelam}, {Roueff},
  {Bachiller}, {Spezzano}, {Rivi{\`e}re-Marichalar},
  {Mart{\'\i}n-Dom{\'e}nech}, \& {Mu{\~n}oz Caro}}]{Esplugues2022}
{Esplugues}, G., {Fuente}, A., {Navarro-Almaida}, D., {et~al.} 2022, \aap, 662,
  A52

\bibitem[{{Ferrante} {et~al.}(2008){Ferrante}, {Moore}, {Spiliotis}, \&
  {Hudson}}]{Ferrante2008}
{Ferrante}, R.~F., {Moore}, M.~H., {Spiliotis}, M.~M., \& {Hudson}, R.~L. 2008,
  \apj, 684, 1210

\bibitem[{Flower \& Launay(1977)}]{Flower1977}
Flower, D.~R. \& Launay, J.~M. 1977, Journal of Physics B: Atomic and Molecular
  Physics, 10, 3673

\bibitem[{{Fuente} {et~al.}(2016){Fuente}, {Cernicharo}, {Roueff}, {Gerin},
  {Pety}, {Marcelino}, {Bachiller}, {Lefloch}, {Roncero}, \&
  {Aguado}}]{Fuente2016}
{Fuente}, A., {Cernicharo}, J., {Roueff}, E., {et~al.} 2016, \aap, 593, A94

\bibitem[{{Fuente} {et~al.}(2019){Fuente}, {Navarro}, {Caselli}, {Gerin},
  {Kramer}, {Roueff}, {Alonso-Albi}, {Bachiller}, {Cazaux}, {Commercon},
  {Friesen}, {Garc{\'\i}a-Burillo}, {Giuliano}, {Goicoechea}, {Gratier},
  {Hacar}, {Jim{\'e}nez-Serra}, {Kirk}, {Lattanzi}, {Loison}, {Malinen},
  {Marcelino}, {Mart{\'\i}n-Dom{\'e}nech}, {Mu{\~n}oz-Caro}, {Pineda},
  {Tafalla}, {Tercero}, {Ward-Thompson}, {Trevi{\~n}o-Morales},
  {Rivi{\'e}re-Marichalar}, {Roncero}, {Vidal}, \& {Ballester}}]{Fuente2019}
{Fuente}, A., {Navarro}, D.~G., {Caselli}, P., {et~al.} 2019, \aap, 624, A105

\bibitem[{{Fuente} {et~al.}(2023){Fuente}, {Rivi{\`e}re-Marichalar},
  {Beitia-Antero}, {Caselli}, {Wakelam}, {Esplugues}, {Rodr{\'\i}guez-Baras},
  {Navarro-Almaida}, {Gerin}, {Kramer}, {Bachiller}, {Goicoechea},
  {Jim{\'e}nez-Serra}, {Loison}, {Ivlev}, {Mart{\'\i}n-Dom{\'e}nech},
  {Spezzano}, {Roncero}, {Mu{\~n}oz-Caro}, {Cazaux}, \&
  {Marcelino}}]{Fuente2023}
{Fuente}, A., {Rivi{\`e}re-Marichalar}, P., {Beitia-Antero}, L., {et~al.} 2023,
  \aap, 670, A114

\bibitem[{{Fuentetaja} {et~al.}(2022){Fuentetaja}, {Ag{\'u}ndez}, {Cabezas},
  {Tercero}, {Marcelino}, {Pardo}, {de Vicente}, \&
  {Cernicharo}}]{Fuentetaja2022}
{Fuentetaja}, R., {Ag{\'u}ndez}, M., {Cabezas}, C., {et~al.} 2022, \aap, 667,
  L4

\bibitem[{{Garrod} {et~al.}(2007){Garrod}, {Wakelam}, \& {Herbst}}]{Garrod2007}
{Garrod}, R.~T., {Wakelam}, V., \& {Herbst}, E. 2007, \aap, 467, 1103

\bibitem[{Georgievskii \& Klippenstein(2005)}]{Georgievskii-Klippenstein:05}
Georgievskii, Y. \& Klippenstein, S.~J. 2005, J.Chem. Phys., 122, 194103

\bibitem[{{Gerner} {et~al.}(2014){Gerner}, {Beuther}, {Semenov}, {Linz},
  {Vasyunina}, {Bihr}, {Shirley}, \& {Henning}}]{Gerner2014}
{Gerner}, T., {Beuther}, H., {Semenov}, D., {et~al.} 2014, \aap, 563, A97

\bibitem[{Goicoechea {et~al.}(2021)Goicoechea, Aguado, Cuadrado, Roncero, Pety,
  Bron, Fuente, Riquelme, Chapillon, Herrera, \& Duran}]{Goicoechea-etal:21}
Goicoechea, J.~R., Aguado, A., Cuadrado, S., {et~al.} 2021, AA, 647, A10

\bibitem[{{Goicoechea} \& {Cuadrado}(2021)}]{Goicoechea2021}
{Goicoechea}, J.~R. \& {Cuadrado}, S. 2021, \aap, 647, L7

\bibitem[{{Goicoechea} {et~al.}(2006){Goicoechea}, {Pety}, {Gerin}, {Teyssier},
  {Roueff}, {Hily-Blant}, \& {Baek}}]{Goicoechea:2006}
{Goicoechea}, J.~R., {Pety}, J., {Gerin}, M., {et~al.} 2006, \aap, 456, 565

\bibitem[{G{\'o}mez-Carrasco {et~al.}(2022)G{\'o}mez-Carrasco,
  F{\'e}lix-Gonz{\'a}lez, Aguado, \& Roncero}]{Gomez-Carrasco-etal:22}
G{\'o}mez-Carrasco, S., F{\'e}lix-Gonz{\'a}lez, D., Aguado, A., \& Roncero, O.
  2022, J. Chem. Phys., 157, 084301

\bibitem[{{Gratier} {et~al.}(2016){Gratier}, {Majumdar}, {Ohishi}, {Roueff},
  {Loison}, {Hickson}, \& {Wakelam}}]{Gratier2016}
{Gratier}, P., {Majumdar}, L., {Ohishi}, M., {et~al.} 2016, \apjs, 225, 25

\bibitem[{Grozdanov \& Solov'ev(1982)}]{Grozdanov-Solovev:82}
Grozdanov, T.~P. \& Solov'ev, E.~A. 1982, J. Phys. B, 15, 1195

\bibitem[{Gu {et~al.}(2006)Gu, Guo, Zhang, Mebel, \& Kaiser}]{GU006:245}
Gu, X., Guo, Y., Zhang, F., Mebel, A.~M., \& Kaiser, R.~I. 2006, Faraday
  Discuss., 245

\bibitem[{Hapka {et~al.}(2012)Hapka, {\.{Z}}uchowski, Szcz{\c{e}}{\'{s}}niak,
  \& Cha{\l}asi{\'{n}}ski}]{Hapka2012}
Hapka, M., {\.{Z}}uchowski, P.~S., Szcz{\c{e}}{\'{s}}niak, M.~M., \&
  Cha{\l}asi{\'{n}}ski, G. 2012, J. Chem. Phys., 137, 164104

\bibitem[{{Hasegawa} \& {Herbst}(1993)}]{Hasegawa1993}
{Hasegawa}, T.~I. \& {Herbst}, E. 1993, \mnras, 261, 83

\bibitem[{Herzberg(1950)}]{Herzberg-diatomics}
Herzberg, G. 1950, Molecular spectra and molecular structure. I. Spectra of
  diatomic molecules (van Nostrand Reinhold Co. (New York))

\bibitem[{Hily-Blant {et~al.}(2022)Hily-Blant, des For{\^e}ts, Faure, \&
  Lique}]{Hily-Blant-etal:22}
Hily-Blant, P., des For{\^e}ts, G.~P., Faure, A., \& Lique, F. 2022, \aap, 658,
  A168

\bibitem[{Huber \& Herzberg(1979)}]{Herzberg-etal:79}
Huber, K.~P. \& Herzberg, G. 1979, Molecular Spectra and Molecular Structure.
  Vol IV. Constants of Diatomic Molecules (Van Nostrand, Toronto)

\bibitem[{{Jim{\'e}nez-Escobar} \& {Mu{\~n}oz Caro}(2011)}]{Jimenez2011}
{Jim{\'e}nez-Escobar}, A. \& {Mu{\~n}oz Caro}, G.~M. 2011, \aap, 536, A91

\bibitem[{{Jim{\'e}nez-Escobar} {et~al.}(2014){Jim{\'e}nez-Escobar}, {Mu{\~n}oz
  Caro}, \& {Chen}}]{Jimenez2014}
{Jim{\'e}nez-Escobar}, A., {Mu{\~n}oz Caro}, G.~M., \& {Chen}, Y.~J. 2014,
  \mnras, 443, 343

\bibitem[{Karplus {et~al.}(1965)Karplus, Porter, \& Sharma}]{Karplus-etal:65}
Karplus, M., Porter, R.~N., \& Sharma, R.~D. 1965, J. Chem. Phys., 43, 3259

\bibitem[{Kendall {et~al.}(1992)Kendall, Dunning, \& Harrison}]{KEN92:6796}
Kendall, R.~A., Dunning, T.~H., \& Harrison, R.~J. 1992, J. Chem. Phys., 96,
  6796

\bibitem[{Klos {et~al.}(2004)Klos, Szczesniak, \& {Chalasinski *}}]{Klos2004}
Klos, J., Szczesniak, M.~M., \& {Chalasinski *}, G. 2004, International Reviews
  in Physical Chemistry, 23, 541

\bibitem[{Konings {et~al.}(2021)Konings, Desrousseaux, Lique, \&
  Loreau}]{Konings-etal:21}
Konings, M., Desrousseaux, B., Lique, F., \& Loreau, J. 2021, J. Chem. Phys.,
  155, 104302

\bibitem[{Laas \& Caselli(2019)}]{Laas-Caselli:19}
Laas, J.~C. \& Caselli, P. 2019, \aap, 624, A108

\bibitem[{{Lucas} \& {Liszt}(2002)}]{LucasLiszt2002}
{Lucas}, R. \& {Liszt}, H.~S. 2002, \aap, 384, 1054

\bibitem[{{McClure} {et~al.}(2023){McClure}, {Rocha}, {Pontoppidan}, {Crouzet},
  {Chu}, {Dartois}, {Lamberts}, {Noble}, {Pendleton}, {Perotti}, {Qasim},
  {Rachid}, {Smith}, {Sun}, {Beck}, {Boogert}, {Brown}, {Caselli}, {Charnley},
  {Cuppen}, {Dickinson}, {Drozdovskaya}, {Egami}, {Erkal}, {Fraser}, {Garrod},
  {Harsono}, {Ioppolo}, {Jim{\'e}nez-Serra}, {Jin}, {J{\o}rgensen},
  {Kristensen}, {Lis}, {McCoustra}, {McGuire}, {Melnick}, {{\~A}-berg},
  {Palumbo}, {Shimonishi}, {Sturm}, {van Dishoeck}, \&
  {Linnartz}}]{McClure2023}
{McClure}, M.~K., {Rocha}, W.~R.~M., {Pontoppidan}, K.~M., {et~al.} 2023,
  Nature Astronomy, 7, 431

\bibitem[{{Minissale} {et~al.}(2016){Minissale}, {Dulieu}, {Cazaux}, \&
  {Hocuk}}]{Minissale2016}
{Minissale}, M., {Dulieu}, F., {Cazaux}, S., \& {Hocuk}, S. 2016, \aap, 585,
  A24

\bibitem[{Nagy \& Lendvay(2017)}]{Nagy-Lendvay:17}
Nagy, T. \& Lendvay, G. 2017, J. Phys. Chem. Lett., 8, 4621

\bibitem[{{Navarro-Almaida} {et~al.}(2023){Navarro-Almaida}, {Bop}, {Lique},
  {Esplugues}, {Rodr{\'\i}guez-Baras}, {Kramer}, {Romero}, {Fuente}, {Caselli},
  {Rivi{\`e}re-Marichalar}, {Kirk}, {Chac{\'o}n-Tanarro}, {Roueff},
  {Mroczkowski}, {Bhandarkar}, {Devlin}, {Dicker}, {Lowe}, {Mason}, {Sarazin},
  \& {Sievers}}]{Navarro2023}
{Navarro-Almaida}, D., {Bop}, C.~T., {Lique}, F., {et~al.} 2023, \aap, 670,
  A110

\bibitem[{{Navarro-Almaida} {et~al.}(2020){Navarro-Almaida}, {Le Gal},
  {Fuente}, {Rivi{\`e}re-Marichalar}, {Wakelam}, {Cazaux}, {Caselli}, {Laas},
  {Alonso-Albi}, {Loison}, {Gerin}, {Kramer}, {Roueff}, {Bachiller},
  {Commer{\c{c}}on}, {Friesen}, {Garc{\'\i}a-Burillo}, {Goicoechea},
  {Giuliano}, {Jim{\'e}nez-Serra}, {Kirk}, {Lattanzi}, {Malinen}, {Marcelino},
  {Mart{\'\i}n-Dom{\`e}nech}, {Mu{\~n}oz Caro}, {Pineda}, {Tercero},
  {Trevi{\~n}o-Morales}, {Roncero}, {Hacar}, {Tafalla}, \&
  {Ward-Thompson}}]{Navarro2020}
{Navarro-Almaida}, D., {Le Gal}, R., {Fuente}, A., {et~al.} 2020, \aap, 637,
  A39

\bibitem[{Oca{\~n}a {et~al.}(2017)Oca{\~n}a, Jim\'enez, Ballesteros, Canosa,
  Anti{\~n}olo, Albadalejo, Ag\'undez, Cernicharo, Zanchet, del Mazo, Roncero,
  \& Aguado}]{Ocana-etal:17}
Oca{\~n}a, A.~J., Jim\'enez, E., Ballesteros, B., {et~al.} 2017, AstroPhys. J.,
  850, 28

\bibitem[{{Palumbo} {et~al.}(1997){Palumbo}, {Geballe}, \&
  {Tielens}}]{Palumbo1997}
{Palumbo}, M.~E., {Geballe}, T.~R., \& {Tielens}, A.~G.~G.~M. 1997, \apj, 479,
  839

\bibitem[{{Palumbo} {et~al.}(1995){Palumbo}, {Tielens}, \&
  {Tokunaga}}]{Palumbo1995}
{Palumbo}, M.~E., {Tielens}, A.~G.~G.~M., \& {Tokunaga}, A.~T. 1995, \apj, 449,
  674

\bibitem[{Páramo {et~al.}(2008)Páramo, Canosa, Le~Picard, \&
  Sims}]{PAR008:9591}
Páramo, A., Canosa, A., Le~Picard, S.~D., \& Sims, I.~R. 2008, J. Phys. Chem.
  A, 112, 9591

\bibitem[{Qu \& Bowman(2016)}]{Qu-Bowman:16}
Qu, C. \& Bowman, J.~M. 2016, J. Phys. Chem. A, 120, 4988

\bibitem[{Quack \& Troe(1974)}]{Quack-Troe:74}
Quack, M. \& Troe, J. 1974, Ber. Bunsenges. Phys. Chem, 78, 240

\bibitem[{Reddy {et~al.}(2003)Reddy, Nazeer~Ahammed, Rama~Gopal, \&
  Baba~Basha}]{RED003:419}
Reddy, R.~R., Nazeer~Ahammed, Y., Rama~Gopal, K., \& Baba~Basha, D. 2003,
  Astrophys. Space Sci., 286, 419

\bibitem[{Riaplov {et~al.}(2003)Riaplov, Wyss, Maier, Panten, Chambaud, Rosmus,
  \& Fabian}]{RIA003:15}
Riaplov, E., Wyss, M., Maier, J.~P., {et~al.} 2003, J. Mol. Spectrosc., 222, 15

\bibitem[{{Rivi{\`e}re-Marichalar} {et~al.}(2019){Rivi{\`e}re-Marichalar},
  {Fuente}, {Goicoechea}, {Pety}, {Le Gal}, {Gratier}, {Guzm{\'a}n}, {Roueff},
  {Loison}, {Wakelam}, \& {Gerin}}]{Riviere2019}
{Rivi{\`e}re-Marichalar}, P., {Fuente}, A., {Goicoechea}, J.~R., {et~al.} 2019,
  \aap, 628, A16

\bibitem[{{Rodr{\'\i}guez-Baras} {et~al.}(2021){Rodr{\'\i}guez-Baras},
  {Fuente}, {Rivi{\'e}re-Marichalar}, {Navarro-Almaida}, {Caselli}, {Gerin},
  {Kramer}, {Roueff}, {Wakelam}, {Esplugues}, {Garc{\'\i}a-Burillo}, {Le Gal},
  {Spezzano}, {Alonso-Albi}, {Bachiller}, {Cazaux}, {Commercon}, {Goicoechea},
  {Loison}, {Trevi{\~n}o-Morales}, {Roncero}, {Jim{\'e}nez-Serra}, {Laas},
  {Hacar}, {Kirk}, {Lattanzi}, {Mart{\'\i}n-Dom{\'e}nech}, {Mu{\~n}oz-Caro},
  {Pineda}, {Tercero}, {Ward-Thompson}, {Tafalla}, {Marcelino}, {Malinen},
  {Friesen}, \& {Giuliano}}]{Rodriguez-Baras2021}
{Rodr{\'\i}guez-Baras}, M., {Fuente}, A., {Rivi{\'e}re-Marichalar}, P.,
  {et~al.} 2021, \aap, 648, A120

\bibitem[{Roncero {et~al.}(2018)Roncero, Zanchet, \& Aguado}]{Roncero-etal:18}
Roncero, O., Zanchet, A., \& Aguado, A. 2018, Phys. Chem. Chem. Phys., 20,
  25951

\bibitem[{{Ruaud} {et~al.}(2016){Ruaud}, {Wakelam}, \& {Hersant}}]{Ruaud2016}
{Ruaud}, M., {Wakelam}, V., \& {Hersant}, F. 2016, \mnras, 459, 3756

\bibitem[{Saito {et~al.}(1987)Saito, Kawaguchi, Yamamoto, Ohishi, Suzuki, \&
  Kaifu}]{SAI87:L115}
Saito, S., Kawaguchi, K., Yamamoto, S., {et~al.} 1987, ApJL, 317, L115

\bibitem[{Sanz-Sanz {et~al.}(2015)Sanz-Sanz, Aguado, Roncero, \&
  Naumkin}]{Sanz-Sanz-etal:15}
Sanz-Sanz, C., Aguado, A., Roncero, O., \& Naumkin, F. 2015, J. Chem. Phys.,
  143, 234303

\bibitem[{Senekowitsch {et~al.}(1990)Senekowitsch, Carter, Rosmus, \&
  Werner}]{Senekowitsch-etal:90}
Senekowitsch, J., Carter, S., Rosmus, P., \& Werner, H.-J. 1990, Chem. Phys.,
  147, 281

\bibitem[{Shingledecker {et~al.}(2020)Shingledecker, Lambers, Laas, Vasyunin,
  Herbst, K{\"a}stner, \& Caselli}]{Shingledecker2020}
Shingledecker, C.~N., Lambers, T., Laas, J.~C., {et~al.} 2020, AstroPhys. J.,
  888, 52

\bibitem[{{Simon} {et~al.}(1997){Simon}, {Stutzki}, {Sternberg}, \&
  {Winnewisser}}]{Simon1997}
{Simon}, R., {Stutzki}, J., {Sternberg}, A., \& {Winnewisser}, G. 1997, \aap,
  327, L9

\bibitem[{Song {et~al.}(2016)Song, Zhang, Gao, \& Meng}]{Song-etal:16}
Song, Y.-Z., Zhang, L.-L., Gao, S.-B., \& Meng, Q.-T. 2016, Scientific Rep., 6,
  37734

\bibitem[{{Spezzano} {et~al.}(2022){Spezzano}, {Fuente}, {Caselli}, {Vasyunin},
  {Navarro-Almaida}, {Rodr{\'\i}guez-Baras}, {Punanova}, {Vastel}, \&
  {Wakelam}}]{Spezzano2022}
{Spezzano}, S., {Fuente}, A., {Caselli}, P., {et~al.} 2022, \aap, 657, A10

\bibitem[{{Sternberg} \& {Dalgarno}(1995)}]{SternbergDalgarno1995}
{Sternberg}, A. \& {Dalgarno}, A. 1995, \apjs, 99, 565

\bibitem[{Stoecklin {et~al.}(1988)Stoecklin, Halvick, \&
  Rayez}]{Stoecklin-etal:88}
Stoecklin, T., Halvick, P., \& Rayez, J.~C. 1988, J. Mol. Struct. (Theochem),
  163, 267

\bibitem[{Stoecklin {et~al.}(1990{\natexlab{a}})Stoecklin, Rayez, \&
  Duguay}]{Stoecklin-etal:90a}
Stoecklin, T., Rayez, J.~C., \& Duguay, B. 1990{\natexlab{a}}, Chem. Phys.,
  148, 381

\bibitem[{Stoecklin {et~al.}(1990{\natexlab{b}})Stoecklin, Rayez, \&
  Duguay}]{Stoecklin-etal:90b}
Stoecklin, T., Rayez, J.~C., \& Duguay, B. 1990{\natexlab{b}}, Chem. Phys.,
  148, 399

\bibitem[{Tarroni {et~al.}(2007)Tarroni, Carter, \& Handy}]{TAR007:1129}
Tarroni, R., Carter, S., \& Handy, N.~C. 2007, Mol. Phys., 105, 1129

\bibitem[{{Vastel} {et~al.}(2018){Vastel}, {Qu{\'e}nard}, {Le Gal}, {Wakelam},
  {Andrianasolo}, {Caselli}, {Vidal}, {Ceccarelli}, {Lefloch}, \&
  {Bachiller}}]{Vastel2018}
{Vastel}, C., {Qu{\'e}nard}, D., {Le Gal}, R., {et~al.} 2018, \mnras, 478, 5514

\bibitem[{Vidal {et~al.}(2017)Vidal, Loison, Jaziri, Ruaud, Gratier, \&
  Wakelan}]{Vidal-etal:17}
Vidal, T. H.~G., Loison, J.-C., Jaziri, A.~Y., {et~al.} 2017, Mon. Not. R.
  Astron. Soc., 469, 435

\bibitem[{Visser {et~al.}(2019)Visser, Beck, Bornhauser, Knopp, van Bokhoven,
  Radi, Gourlaouen, \& Marquardt}]{VIS019:1645}
Visser, B., Beck, M., Bornhauser, P., {et~al.} 2019, Mol. Phys., 117, 1645

\bibitem[{Voronin(2004)}]{Voronin:02}
Voronin, A.~I. 2004, Chem. Phys., 297, 49

\bibitem[{{Wakelam} {et~al.}(2021){Wakelam}, {Dartois}, {Chabot}, {Spezzano},
  {Navarro-Almaida}, {Loison}, \& {Fuente}}]{Wakelam2021}
{Wakelam}, V., {Dartois}, E., {Chabot}, M., {et~al.} 2021, \aap, 652, A63

\bibitem[{Werner \& Knowles(1988{\natexlab{a}})}]{Werner-Knowles:88}
Werner, H.~J. \& Knowles, P.~J. 1988{\natexlab{a}}, J. Chem. Phys., 89, 5803

\bibitem[{Werner \& Knowles(1988{\natexlab{b}})}]{Werner-Knowles:88b}
Werner, H.~J. \& Knowles, P.~J. 1988{\natexlab{b}}, Chem. Phys. Lett., 145, 514

\bibitem[{Werner {et~al.}(2012)Werner, Knowles, Knizia, Manby, \&
  Sch{\"u}tz}]{MOLPRO-WIREs}
Werner, H.-J., Knowles, P.~J., Knizia, G., Manby, F.~R., \& Sch{\"u}tz, M.
  2012, WIREs Comput Mol Sci, 2, 242

\bibitem[{Woon \& Herbst(2009)}]{Woon-Herbst:09}
Woon, D.~E. \& Herbst, E. 2009, ApJ. Sup. Series, 185, 273

\bibitem[{Zanchet {et~al.}(2018)Zanchet, del Mazo, Aguado, Roncero, Jim\'enez,
  Canosa, Ag\'undez, \& Cernicharo}]{Zanchet-etal:18}
Zanchet, A., del Mazo, P., Aguado, A., {et~al.} 2018, PCCP, 20, 5415

\bibitem[{Zanchet {et~al.}(2016)Zanchet, Roncero, \& Bulut}]{Zanchet-etal:16}
Zanchet, A., Roncero, O., \& Bulut, N. 2016, Phys. Chem. Chem. Phys., 18, 11391

\bibitem[{Zanchet {et~al.}(2009)Zanchet, Roncero, Gonz{\'a}lez-Lezana,
  Rodr{\'\i}guez-L{\'o}pez, Aguado, Sanz-Sanz, \&
  G{\'o}mez-Carrasco}]{Zanchet-etal:09b}
Zanchet, A., Roncero, O., Gonz{\'a}lez-Lezana, T., {et~al.} 2009, J. Phys.
  Chem. A, 113, 14488

\bibitem[{Zeimen {et~al.}(2003)Zeimen, Klos, Groenenboom, \& van~der
  Avoird}]{Zeimen-etal:03}
Zeimen, W.~B., Klos, J., Groenenboom, G.~C., \& van~der Avoird, A. 2003, J.
  Chem. Phys., 118, 7340

\bibitem[{Zhang {et~al.}(2018)Zhang, Song, Gao, \& Meng}]{Zhang-etal:18b}
Zhang, L.~L., Song, Y.~Z., Gao, S.~B., \& Meng, Q.~T. 2018, J. Phys. Chem. A,
  122, 4390

\end{thebibliography}

\end{document}